\documentclass[twocolumn]{aastex631}
\usepackage[utf8]{inputenc}
\usepackage{amsmath,amsxtra,amssymb,latexsym,amscd,amsthm}
\usepackage{mathrsfs}
\usepackage[mathscr]{eucal}
\usepackage{mathrsfs}
\usepackage{makeidx}
\usepackage{soul}
\usepackage{graphicx}
\usepackage{subfigure}
\usepackage{booktabs}
\usepackage{multirow}
\usepackage{threeparttable}
\usepackage{hyperref}
\usepackage[T1]{fontenc}

\newcommand{\mbs}[1]{\boldsymbol{#1}}

\newcommand{\tx}[1]{\text{#1}}
\newcommand{\pt}{\partial}

\shorttitle{Dust Trapping}
\graphicspath{{./}{figures/}}

\begin{document}

\title{Leaky Dust Traps in Planet-Embedded Protoplanetary Disks}

\author[0000-0002-7575-3176]{Pinghui Huang}
\affiliation{CAS Key Laboratory of Planetary Sciences, Purple Mountain Observatory, Chinese Academy of Sciences, Nanjing 210008, People’s Republic of China}
\affiliation{Department of Physics \& Astronomy, University of Victoria, Victoria, British Columbia, V8P 5C2, Canada}

\author[0009-0004-6973-3955]{Fangyuan Yu}
\affiliation{Zhiyuan College, Shanghai Jiao Tong University, Shanghai 200240, People’s Republic of China; yufangyuan@sjtu.edu.cn}
\affiliation{Tsung-Dao Lee Institute, Shanghai Jiao Tong University, Shanghai 201210, People’s Republic of China}

\author[0000-0002-1228-9820]{Eve J.~Lee}
\affiliation{Department of Astronomy \& Astrophysics, University of California, San Diego, La Jolla, CA 92093-0424, USA; evelee@ucsd.edu}
\affiliation{Department of Physics and Trottier Space Institute, McGill University, 3600 rue University, H3A 2T8 Montreal QC, Canada}
\affiliation{Trottier Institute for Research on Exoplanets (iREx), Universit\'e de Montr\'eal, Canada}

\author[0000-0001-9290-7846]{Ruobing Dong}
\affiliation{Kavli Institute for Astronomy and Astrophysics, Peking University, Beijing 100871, People’s Republic of China; rbdong@pku.edu.cn}
\affiliation{Department of Physics \& Astronomy, University of Victoria, Victoria, British Columbia, V8P 5C2, Canada}

\author[0000-0001-6906-9549]{Xue-Ning Bai}
\affiliation{Institute for Advanced Study, Tsinghua University, Beijing 100084, People’s Republic of China; xbai@mail.tsinghua.edu.cn}
\affiliation{Department of Astronomy, Tsinghua University, Beijing 100084, People’s Republic of China}

\begin{abstract}

From the survival of dust disks for a few Myr to the establishment of chemical dichotomy, dust traps are expected to play a pivotal role in sculpting the protoplanetary disks and the early planet formation process. These traps may not be perfect as evidenced by both numerical simulations and the observations of disks with gaps and cavities inside which we detect some amount of both gas and dust. Using two-fluid hydrodynamic global simulations in both two-dimensions (2D) and three-dimensions (3D), we directly compute the dynamics of dust grains as they aerodynamically interact with the disk gas that is being perturbed by an embedded planet. In both 2D and 3D, we find the dust trap to be more leaky for lower mass planet and for higher turbulent $\alpha$. More crucially, we find the fraction of the dust mass that remain trapped within the pressure bump can be up to an order of magnitude more reduced in 3D compared to 2D with all else equal. Our simulations show a complex behavior of dust radial motion that is both azimuthally and poloidally non-uniform, with the overall dynamics dominated by the dust coupling to the gas flow even for Stokes number 0.1. The leaky traps we find suggest pebble isolation mass is likely not truly isolating and that gap-opening planets do not establish an unconditional impermeable barrier. Our findings have implications for recent JWST MINDS results, which show that volatiles, including water, are present in the inner regions of disks hosting outer dust rings.

\end{abstract}

\section{Introduction}~\label{sec:intro}

Ring-like substructures are ubiquitous in radio interferometric images of nearby bright protoplanetary disks 
\citep[e.g.,][]{Andrews18,Long19}.
These local concentrations of dust are often explained as trapping within a gas pressure maximum whereby the aerodynamic drag collects the particles towards the center of the maximum \citep[e.g.,][]{Pinilla12,deJuanOvelar2016, Pinilla2021, Jiang2024}.
One of the major sources of this gas pressure maximum is the tidal perturbation by an embedded planet \citep[e.g.,][]{Goldreich80,Ogilvie02}. Within the uncertainties of the particle Stokes number, defined as the particle aerodynamic stopping time in units of local orbital time, and the strength of turbulence in gas, a wide range of planet masses and multiplicity can be invoked to explain many of the observed rings~\citep{Dong15,DongLi2017,Bae2017}. 

Irrespective of the sources, the existence of dust traps have fundamental implications on the evolution of protoplanetary disks and the growth of the planets. It solves the classical problem that dust grains 
marginally coupled to the gas would undergo extremely rapid radial drift, arriving at the inner edge of the gas disk within 100s of years, inconsistent with the detection of such dust at large distances from the star in both SED and in imaging observations \citep[e.g.,][]{Andrews18,Birnstiel24}. Trapping the dust at the gap edges gives a natural solution to this conundrum. Yet another implication of dust traps can be found in the concept of pebble isolation masses~\citep{Johansen2019,Bitsch2019}, whereby planets of sufficiently high mass carves out a gap exterior to the planet's orbit. Incoming pebbles (particles with Stokes number $\lesssim$1) are trapped at the edge of this gap and the influx of pebble stops, halting further mass growth of the planet and starving the region interior to the massive planet of solids. The latter concept in particular has been used to argue for the relative lack of inner super-Earths for systems that harbor an outer giant~\citep{Bonomo2017,Mulders2020} although recent observations show otherwise~\citep[e.g.,][]{ZhuWu2018,Bryan19,BryanLee2024}.

The efficiency at which the grains and pebbles are trapped, however, is not certain. Early simulations found the traps to be imperfect and that particles of small Stokes number can be readily filtered through, owing to their strong coupling to the gas advective motion \citep[e.g.,][]{Zhu12}. The reduced trapping efficiency at small particle sizes is consistent with the observations of transitional disks~\citep{Dong2012,Follette2015} including PDS 70 \citep{Pinilla24}. Using two-dimensional local shearing box simulations with an embedded planet, \citet{Lee22} found that the trapping efficiency, defined as the ratio of the total sum of the dust mass trapped over the total sum of the dust mass drifted in, was close to one but not exactly one, suggesting that the traps are indeed imperfect---and at Stokes number $\lesssim$0.05, even found the ring to be destroyed over time as the dust grains are advected away with the gas flow. 
The real world is three-dimensional however and planetary perturbations are necessarily non-axisymmetric. Hydrodynamic processes are known to show different behaviors between 2D and 3D simulations~\citep{Fung2015,FungChiang2016}. In this paper, we aim to investigate in detail the dust dynamics in the vicinity of planet-driven pressure bumps in global, three-dimensional simulations, elevating prior calculations to better physical realism.

Our paper is organized as follows. We describe our numerical methods in Section \ref{sec:method}. Results are presented in Section \ref{sec:results} comparing 2D and 3D. In Section \ref{sec:discussion}, we analyze in detail the morphology of 3D dust motion and discuss their implication on planet formation. We summarize and conclude in Section \ref{sec:conclusion}.

\section{Numerical Methods}~\label{sec:method}

We conduct direct numerical simulations of gas and dust using the multifluid dust module within Athena++, a shock-capturing magnetohydrodynamics code implemented using the finite volume method~\citep{Stone2020}. In our simulations, we treat dust as another neutral and pressureless fluid, similar to FARGO3D~\citep{Benitez2019FARGO3D} and MPI-AMRVAC~\citep{Keppens2023MPIAMRVAC}. It is important to note that the multifluid approximation is applicable to dust with small Stokes number, $St < 1$---which is satisfied in our simulations---in contrast to other codes implemented by the particle-mesh method, Pencil~\citep{JohansenYoudin2007}, Athena~\citep{Bai2010particle}, PLUTO~\citep{MignoneFlock2019Dust}. 

The main advantage of our multifluid method is the ease of mimicing turbulence within dust by directly simulating dust diffusion which is non-trivial for Lagrangian dust particles. Further details on the multifluid module of Athena++ can be found in~\cite{HuangBai2022}.

\subsection{General Equations and Numerical Setups}~\label{subsec:equationsAndsetup}

We solve the following equations to track the motion of gas and dust:

\begin{equation}
\frac{\pt \rho_{\tx{g}}}{\pt t} + \nabla \cdot \left(\rho_{\tx{g}} \mbs{v}_{\tx{g}}\right) =0\ ,
\label{eq:gas_con}
\end{equation}

\begin{equation}
\begin{aligned}
  \frac{\pt \left(\rho_{\tx{g}}\mbs{v}_{\tx{g}}\right)}{\pt t} &+  \nabla\cdot\left(\rho_{\tx{g}}\mbs{v}_{\tx{g}} \mbs{v}_{\tx{g}} + P_{\tx{g}}\mathsf{I}+\mbs{\Pi}_{\nu}\right)= \\
  & \rho_{\tx{g}} \nabla \Phi_{*} + \rho_{\tx{g}} \nabla \Phi_{\tx{p}} +\rho_{\tx{d}} \frac{\mbs{v}_{\tx{d}} - \mbs{v}_{\tx{g}}}{St} \;\Omega_\text{K}\ ,
\end{aligned}
\label{eq:gas_mom}
\end{equation}

\begin{equation}
\begin{aligned}
  &\frac{\pt E_{\tx{g}}}{\pt t} + \nabla \cdot\left[\left(E_{\tx{g}}+P_{\tx{g}}\right) \mbs{v}_{\tx{g}}+\mbs{\Pi}_{\nu} \cdot \mbs{v}_{\tx{g}}\right]= \rho_{\tx{g}} \nabla \left(\Phi_* + \Phi_\tx{p}\right) \cdot \mbs{v}_{\tx{g}} \\
  &+ \rho_{\tx{d}} \frac{(\mbs{v}_{\tx{d}} - \mbs{v}_{\tx{g}})\cdot \mbs{v}_{\tx{g}}\Omega_\text{K}}{St}  + \rho_{\tx{d}} \frac{\left(\mbs{v}_{\tx{d}} - \mbs{v}_{\tx{g}}\right)^2 \;\Omega_\text{K}}{St}\ \\
  &- \frac{\rho_\tx{g}}{\gamma - 1} \frac{\left(T - T_\tx{init}\right)\Omega_\tx{K}}{\beta},
\end{aligned}
\label{eq:gas_erg}
\end{equation}

\begin{equation}
\frac{\pt \rho_{\tx{d}}}{\pt t} +\nabla \cdot \left[\rho_{\tx{d}} \mbs{v}_{\tx{d}} - \rho_\tx{g} D_\tx{d} \nabla \left(\frac{\rho_\tx{d}}{\rho_\tx{g}}\right)\right] = 0\ ,
\label{eq:dust_con}
\end{equation}

\begin{equation}
\begin{aligned}
  \frac{\pt \rho_{\tx{d}} \left(\mbs{v}_{\tx{d}} + \mbs{v}_{\tx{d,dif}}\right)}{\pt t} &+ \nabla\cdot(\rho_{\tx{d}}\mbs{v}_{\tx{d}} \mbs{v}_{\tx{d}} + \mbs{\Pi}_{\tx{dif}}) = \\
  &\rho_{\tx{d}} \nabla \Phi_{*} + \rho_{\tx{d}} \nabla \Phi_{\tx{p}} + \rho_{\tx{d}}\frac{\mbs{v}_{\tx{g}} - \mbs{v}_{\tx{d}}}{{St}}\;\Omega_\text{K}.
\end{aligned}
\label{eq:dust_mom}
\end{equation}

The top three equations are the continuity, momentum and energy equations for gas. The bottom two equations are the continuity and momentum equations for dust. The subscripts ``d'' and ``g'' denote ``dust'' and ``gas'', respectively. We consider a single dust species characterized by a constant Stokes number, defined as \(\text{St} \equiv T_\text{s} \Omega_\text{K} = 0.1\), where \(T_\text{s}\) and $\Omega_\text{K} \equiv \sqrt{GM_\star/R^3}$ are the dust stopping time and the Keplerian orbital frequency, respectively with $G$ gravitational constant, $M$ the mass of the central star, and $R$ the stellocentric radial distance (in the cylindrical coordinate).
In these equations, $\rho$ represents density, $P_\tx{g}$ represents gas pressure, $v$ represents velocity, $\mathsf{I}$ represents the identity matrix, $\nabla \Phi_*$ represents the central stellar gravity, i.e., $\Phi_* = GM/r$, where $r$ is the radial location from the star in spherical coordinate, and $\nabla \Phi_\tx{p}$ is planetary gravity:

\begin{equation}
\begin{aligned}
  \Phi_\tx{p} = &-\frac{G M_\tx{p}}{\left[R^2+R_\tx{p}^2-2 R R_\tx{p} \cos \left(\phi-\Omega_\tx{p} t\right)+r_\tx{s}^2\right]^{1 / 2}}  \\
  &+ \frac{G M_\tx{p} R \cos \left(\phi-\Omega_\text{p} t\right)}{R_\tx{p}^2} - \frac{GM_\text{p}z}{(z^2+r_\text{s}^2)^{3/2}},
\end{aligned}
\end{equation}
where $R$ is the radial location from the star in the cylindrical coordinate, $\phi$ is the azimuthal angle, $r_\text{s} = 0.6 r_\text{Hill}$ is the gravitational softening factor, $r_\text{Hill} \equiv (M_p/3M_\star)^{1/3}R_p$ is the Hill radius of the planet, $R_{\rm p}$ is the radial location of the planet (set to 1), $\phi_p$ the azimuthal location of the planet, $\Omega_{\rm p}$ is the Keplerian orbital frequency $\Omega_{\rm K} \equiv \sqrt{GM/R^3}$ evaluated at $R=R_{\rm p}$, and $z$ is the vertical height in cylindrical coordinate. The second term on the right-hand side is a correction for the reference frame. We keep the planet fixed at $R_p = 1$ without migration to study with better clarity the dynamics of dust-gas interaction near a pressure bump.

In terms of gas thermodynamics, we fix the adiabatic index to $\gamma= 1.4$, and we select the cooling parameter $\beta = 1$, to suppress the vertical shear instability in our simulations~\citep{LinYoudin2015}.

We incorporate gas viscosity which mimics the effects from external gas turbulence, described by the viscous stress tensor $\mbs{\Pi}_{\nu}$:
\begin{equation}
  \Pi_{\nu,ij}=\rho_{\tx{g}} \nu_\tx{g}\left(\frac{\pt v_{\tx{g},i}}{\pt x_{\tx{g},j}}+\frac{\pt v_{\tx{g},j}}{\pt x_{\tx{g},i}}-\frac{2}{3} \delta_{ij} \nabla \cdot \mbs{v_{\tx{g}}}\right)\ ,
\end{equation}
where $\nu_\tx{g}$ is the kinematic viscosity, $x$ represents the spatial coordinates, and $\delta_{ij}$ is the kronecker-delta. 

For the equations of dust, we explicitly decompose the dust motion into advective $\mbs{v}_{\rm d}$ and diffusive $\mbs{v}_{\rm d,dif}$ components \citep{HuangBai2022}. Following \citet{YoudinLithwick2007}, we express the dust diffusion coefficient:

\begin{equation}
  D_\tx{d} = \frac{\nu_\tx{g}}{1+St^2},
\end{equation}
and we define the dust diffusive tensor $\Pi_\tx{dif}$ as~\citep{HuangBai2022}
\begin{equation}
  \Pi_{\tx{dif},ij}=v_{\tx{d},j} \mathscr{F}_{\tx{dif},i} + v_{\tx{d},i} \mathscr{F}_{\tx{dif},j},
\label{eq:momflx}
\end{equation}
where
\begin{equation}
  \mbs{\mathscr{F}}_{\tx{dif}} \equiv -\rho_{\tx{g}} D_{\tx{d}} \nabla \left(\frac{\rho_{\tx{d}}}{\rho_{\tx{g}}}\right) = \rho_{\tx{d}}\mbs{v}_{\tx{d,dif}}.
  \label{eq:dust_diff}
\end{equation}
We ignore the effects of self-gravity (gas or dust) and the magnetic field in this study, deferring them to future works. 

We utilize the cylindrical $(R,\phi,z)$ and the spherical-polar coordinate $(r, \theta, \phi)$ within Athena++ for conducting 2D ($z = 0$) and 3D numerical simulations, respectively. The relationships between the spherical-polar and cylindrical coordinates are given by: $R = r \sin \theta$, $z = r \cos \theta$, $r = \sqrt {R^2+z^2}$ and $\theta = \arctan{z/R} + \pi/2$.

We employ static mesh refinement (SMR) to perform 2D and 3D simulations. The root grids for the 2D simulations consist of $64 \times 192$ cells in the $R$-$\phi$ plane. The computational domain spans $R \in [0.3, 2.3]$ with an $x1rat \equiv \delta R_{n+1}/\delta R_{n} = 1.002$ ($\delta R$ is the radial width of grid), and $\phi \in [0, 2\pi]$ with a uniform distribution. Two levels of mesh refinement are applied to refine the planetary orbital region $R \in [0.9, 1.4]$. 
We have verified that doubling the resolution for 2D simulations make no significant difference, validating our choice of baseline resolution.

The setup for the 3D simulations is similar to that of the 2D cases. The root grids of the 3D simulations consist of $64\times32\times192$ in $r-\theta-\phi$ space. The computational domain spans $r \in [0.3, 2.3]$ with an $x1rat$ of 1.002, $\theta \in [\pi/2 - 0.25, \pi/2 + 0.25]$ with a uniform distribution and $\phi \in [0, 2\pi]$ with a uniform distribution. Two levels of the finest mesh cover the regions $r \in [0.9, 1.4]$ and $\theta \in [\pi/2 - 0.125, \pi/2 + 0.125]$. 
Visualization of the mesh structure for 2D and 3D simulations is presented in Appendix~\ref{app:mesh}.

\subsection{Initial and Boundary Conditions}~\label{subsec:ICandBC}
Since disks are rotational objects, we set the initial and boundary conditions based on 3D cylindrical coordinate $(R, \phi, z)$, which is more consistent with the rotational nature of disks. 

In the 2D simulations, the initial gas surface density is set as:
\begin{equation}
    \Sigma_\text{g,init,2D} (R) = \Sigma_0 \left(\frac{R}{R_\text{p}}\right)^{-1}\\
\end{equation}
where $\Sigma_0 \equiv 1$. The initial dust surface density is set to
\begin{equation}
    \Sigma_\text{d,init,2D} (R) = \epsilon_\text{init}\Sigma_\text{g,init,2D}
\end{equation}
where $\epsilon_\text{init} \equiv 0.01$ as the initial dust-gas denisty ratio.

In the 3D simulations, following \citet{Nelson2013}, we adopt the initial gas bulk density $\rho_\text{g,init,3D}$ as:
\begin{align}
  \rho_\tx{g,init,3D}(R,z) &= \rho_0 \left(\frac{R}{R_\text{p}}\right)^{-9/4} \nonumber \\
  &\times \exp \left[\frac{GM}{c_\tx{s,init}^2}\left(\frac{1}{\sqrt{R^2+z^2}}-\frac{1}{R}\right) \right],
\end{align}
where $\rho_0 \equiv 1$, so that $\Sigma_\text{g,init,3D} \propto R^{-1}$, and the initial sound speed
\begin{equation}
    c_\tx{s,init}^2(R) = c_\tx{s,0}^2 \left(\frac{R}{R_\text{p}}\right)^{-1/2},
\end{equation}
where $c_\tx{s,0} \equiv 0.05$. 
Like the initial dust surface density in 2D simulations, the initial dust bulk density is set to
\begin{equation}
    \rho_\text{d,init,3D} (R,z) = \epsilon_\text{init}\rho_\text{g,init,3D}.
\end{equation}

We set the gas initial velocity in 3D to
\begin{equation}
  v_{\text{g},R,\text{init,3D}} = -\frac{3\nu}{2R} = -\frac{3\alpha c_\text{s} H_\text{g}}{2R},
\end{equation}
\begin{equation}
  v_{\text{g},\phi,\text{init,3D}} = R \Omega_\tx{K} \left[\frac{1}{2} -\frac{11}{4}\left(\frac{H_\text{g}}{R}\right)^2 + \frac{R}{2\sqrt{R^2+z^2}}\right]^{\frac{1}{2}},
\end{equation}
\begin{equation}
  v_{\text{g},z,\text{init,3D}} = 0,
\end{equation}
where $H_\tx{g}(R) \equiv c_\tx{s}/\Omega_\tx{K}$.
We control the gas viscosity $\nu_g \equiv \alpha c_s H_g$ through the non-dimensional Shakura-Sunyaev $\alpha$. By using a constant $\alpha$, we implicitly assume isotropic turbulence. Since the dust diffusion coefficient $D_d \propto \nu_g$, it follows that the turbulent motion of dust is also assumed to be isotropic.

The initial 3D cylindrical radial, azimuthal and vertical dust velocities are set to the steady state aerodynamic drift rate from \citet{NakagawaSekiyaHayashi1986}:
\begin{equation}
  v_{\tx{d},R,\tx{init,3D}} = \frac{2}{St + \left(1+\epsilon_\text{init}\right)^2 St^{-1}}\eta v_\tx{K},
  \label{eq:3D_dust_v_R}
\end{equation}
\begin{equation}
  v_{\tx{d},\phi,\tx{init,3D}} = \left[ 1 + \frac{\left(1+\epsilon_\text{init}\right) \eta}{\left(1+\epsilon_\text{init}\right)^2+ St^{2}} \right] v_\tx{K},
  \label{eq:3D_dust_v_phi}
\end{equation}
\begin{equation}
  v_{\tx{d},z,\tx{init,3D}} = 0.
  \label{eq:3D_dust_v_z}
\end{equation}
where $\eta \equiv 1/2 (H_\text{g}/R)^2 d \log P/d \log R$ represents the strength of the radial gas pressure gradient.
Based on the 3D initial velocities, the initial velocities of gas and dust in 2D simulations are trivially determined by taking $z = 0$ and $r = R$ in above equations:
\begin{equation}
    \mbs{v}_\text{g,init,2D}(R) = \mbs{v}_\text{g,init,3D}(R,0),
\end{equation}
\begin{equation}
    \mbs{v}_\text{d,init,2D}(R) = \mbs{v}_\text{d,init,3D}(R,0).
\end{equation}

We begin by running 100 planetary orbits of gas-only simulations to allow the gas profile to relax and the gap to form, before introducing dust into the simulations. During the initial 100 orbits, dust is fixed at its initial positions, and the dust back-reaction on the gas is disabled---the gas does not feel the presence of dust. After 100 orbits, the dust is released, and the back-reaction is simultaneously activated.

Periodic boundary conditions are applied in the azimuthal direction for both gas and dust in the 2D and 3D simulations. The radial boundary conditions for gas and dust in both 2D and 3D simulations are set according to their respective initial conditions at the corresponding locations. In the meridional direction of the 3D simulations, the gas density and velocities are configured similarly to those in the radial direction. For dust, the density is set to zero in the ghost cells at the meridional boundaries, and the velocities are initialized to their respective initial values.

All 2D and 3D simulations use the Piecewise Linear Method (PLM) for spatial reconstruction and the Van-Leer 2 (VL2) time integrator, with a Courant-Friedrichs-Lewy (CFL) number of 0.3. Additionally, the orbital advection method~\citep{Masset2000,Masset2002} is applied to the azimuthal velocities of both gas and dust in all simulations to enhance computational efficiency.

\subsection{Parameters to Vary}

Table~\ref{tab:Models} summarizes the 2D and 3D models analyzed in this study. We investigate four 2D and four 3D runs, varying the planetary masses and turbulent viscosity. The planetary mass is also marked in the unit of thermal mass $M_\text{th}$. We choose trans-thermal and sub-thermal planet masses as super-thermal planets often form nearly perfect dust traps already and therefore are not interesting.
In all cases, we adopt a constant thermal cooling parameter $\beta = 1$ in our numerical setup. All the simulations are run up to 1500 orbits. We choose relatively large St and $\alpha$ to optimize against computational cost. In particular, our choice ensures the gap profiles carved by planets reach steady states within $\lesssim$100 orbits and the dust grains that were originally near the outer edge of the disk can reach the dust trap within a similar timeframe. A more extensive parameter study is a subject of future work.

\begin{table}
%\centering
\begin{normalsize}
\caption{List for 2D \& 3D Simulations}
\begin{tabular}{ccccc}
\toprule
\hline
Model Name      & $M_\text{p}/M_*$ ($M_\text{th}$) & $\alpha$ \\
\hline
2D \& 3D Model A         &  $3\times 10^{-4}$ (2.4)   &  $10^{-3}$ \\
2D \& 3D Model B         &  $3\times 10^{-4}$ (2.4)   &  $10^{-2}$ \\
2D \& 3D Model C         &  $1\times 10^{-4}$ (0.8)   &  $10^{-3}$ \\
2D \& 3D Model D         &  $3\times 10^{-5}$ (0.24)  &  $10^{-3}$ \\
\bottomrule
\end{tabular}
\label{tab:Models}
\end{normalsize}
\end{table}

\subsection{Definition of Trapping Efficiency}~\label{subsec:diagnostics}
We define the flux trapping ratio as the net dust mass flux through the dust trap divided by the total mass flux entering the trap:

\begin{equation}
  \epsilon_{\rm flux} (t) =\frac{\dot{M}_{\text{ring,outer}}-\dot{M}_{\text{ring,inner}}}{\dot{M}_{\text{ring,outer}}}\\
  \label{eq:trapratio}
\end{equation}
where the subscript `outer' corresponds to the outer edge and the subscript `inner' to the inner edge of the dust ring. In 3D simulations,
\begin{equation}
\begin{aligned}
&\dot{M}_{\text{ring}} \equiv \dot{M}_{r}(t)\\
\dot{M}_{r}(t) \equiv & \int^{2\pi}_{0} \int^{\theta_\text{max}}_{\theta_\text{min}} \rho_\text{d} \left( v_{\text{d},r} + v_{\text{d},r,\text{dif}} \right) r^2 \sin \theta d\theta d\phi\\
\end{aligned}
\label{eq:flux_ring_3D}
\end{equation}
where $r_{\text{in}}$ and $r_{\text{out}}$ denote the boundaries of the dust trap. We take $\theta_{\rm min}$ and $\theta_{\rm max}$ at our simulation boundaries where the dust vertical flux is zero so we do not include the vertical flux in our definition of $\dot{M}_{\rm ring}$.
In 2D simulations,
\begin{equation}
\begin{aligned}
&\dot{M}_{\text{ring}} \equiv \dot{M}_{R}(t)\\
\dot{M}_{R}(t) \equiv & \int^{2\pi}_{0}  \Sigma_\text{d} \left( v_{\text{d},R} + v_{\text{d},R,\text{dif}} \right) R d\phi.
\end{aligned}
\label{eq:flux_ring_2D}
\end{equation}
Here, we explicitly and separately follow the advective and diffusive velocities of the dust following equations \ref{eq:dust_mom} and \ref{eq:dust_diff}.

\begin{figure*}
\centering
\includegraphics[scale=0.33]{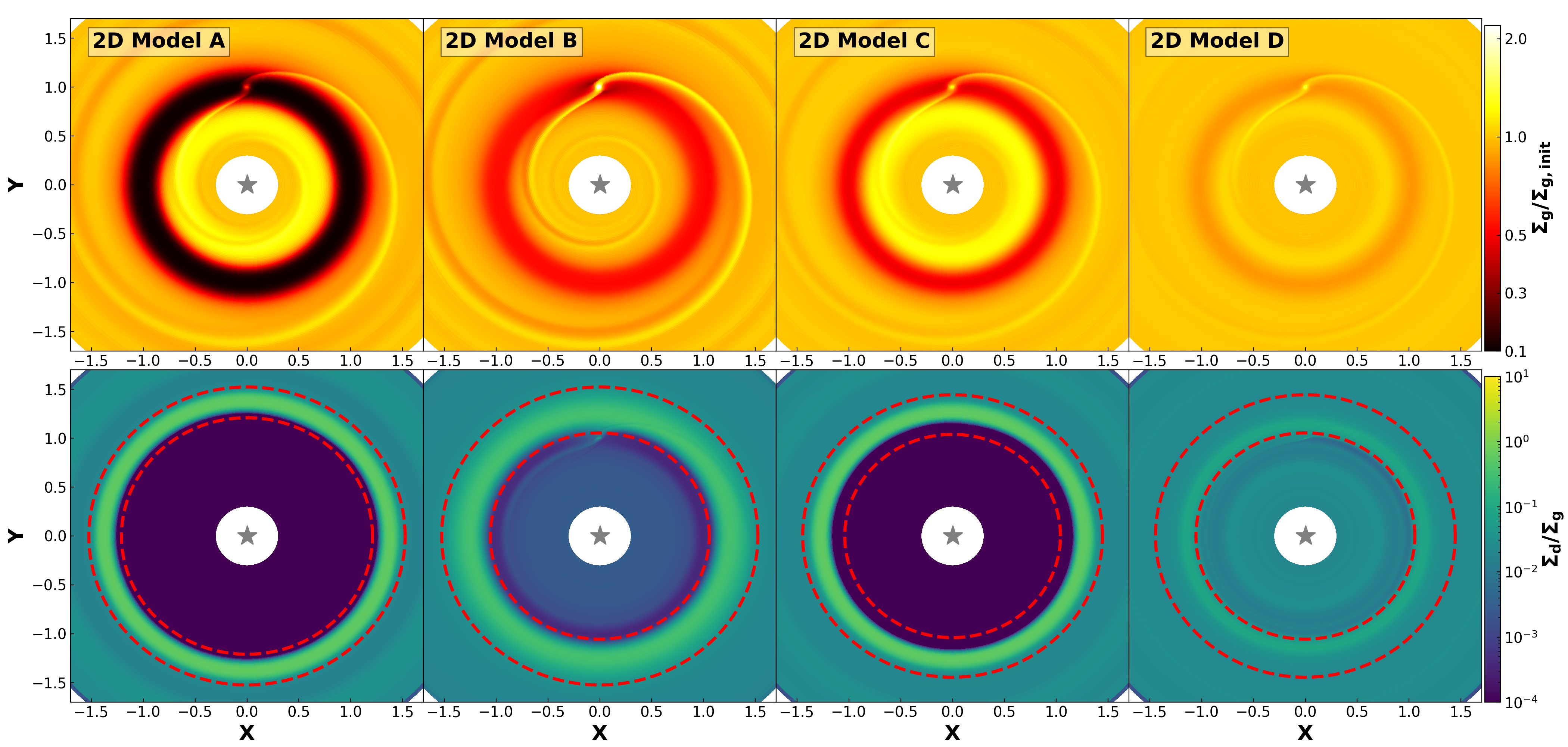}
\caption{The top four panels show the gas surface densities normalized by the initial value, $\Sigma_\text{g}/\Sigma_\text{g,init}$, for the 2D models at \textbf{1500} orbits. The bottom four panels display the dust-to-gas surface density ratios, $\Sigma_\text{d}/\Sigma_\text{g}$, for the same 2D models at 1500 orbits. The x and y axes are normalized by the radial location of planets. Red dashed lines represent the inner boundary and the outer boundary of dusty rings. In Model D, the inner boundary appears slightly outside the visual edge of the dust ring because the edge is already inside $r_{\rm Hill}$. Our $\epsilon_{\rm flux}$ and $\epsilon_{\rm mass}$ for Model D change negligibly when we bring this inner edge down to $r_s$.}
\label{fig:2D_density_ratios}
\end{figure*}

\begin{figure*}
\centering
\includegraphics[scale=0.35]{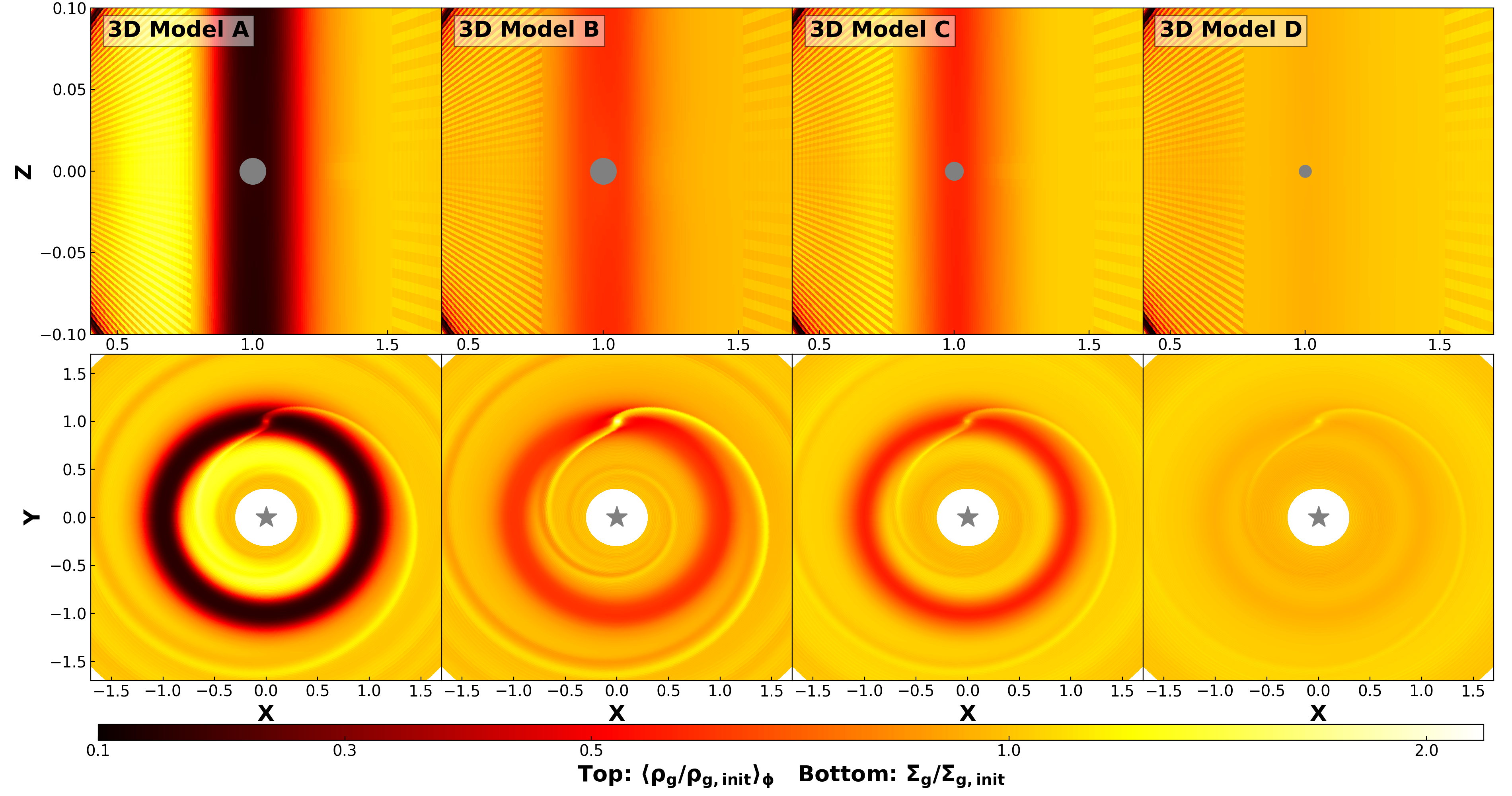}
\caption{
The top four panels show the azimuthally averaged normalized gas densities, $\langle \rho_\text{g}/\rho_\text{g,init} \rangle_\phi$, for the 3D models at 1500 orbits. The bottom four panels display the normalized gas surface densities, $\Sigma_\text{g}/\Sigma_\text{g,init}$, for the same 3D models at 1500 orbits. The size of grey circles are proportional to the Hill radii of planets. The stripes in the upper panels are the plotting artifacts caused by the mesh refinement.}
\label{fig:3D_gas_densities}
\end{figure*}

\begin{figure}[htp]
\centering
\includegraphics[width=1.0\linewidth]{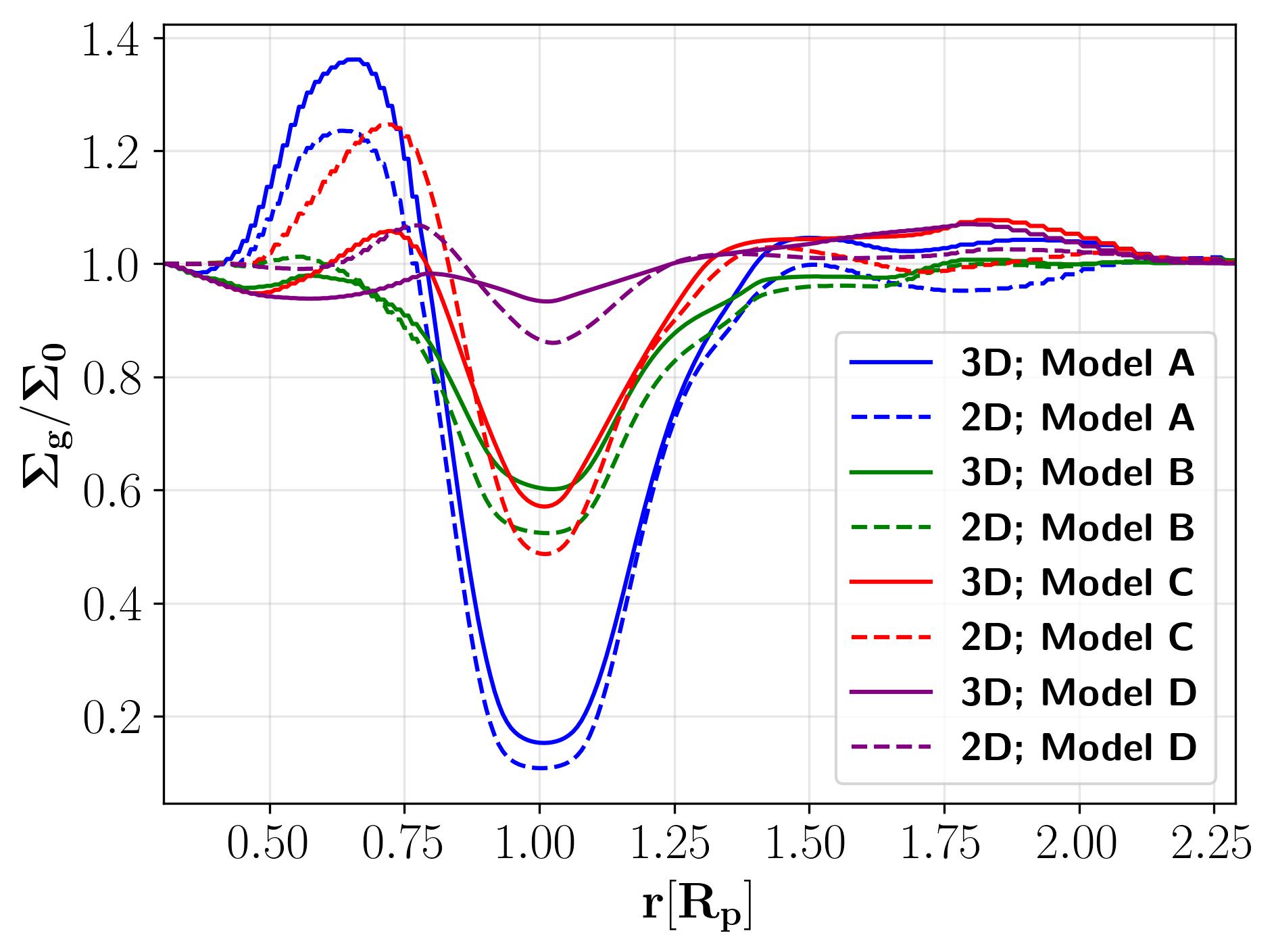}
\caption{
Surface density profiles of the gas, normalized by its initial value and azimuthally averaged excluding a small region around the planet ($\phi_p - \Delta/R_p < \phi < \phi_p + \Delta/R_p$), shown at 1500 planetary orbits. The 2D and 3D results are in close agreement, except for very shallow gaps such as in Model D.}
\label{fig:Sigma_2D_3D}
\end{figure}

\begin{figure*}[htp]
\centering
\includegraphics[scale=0.35]{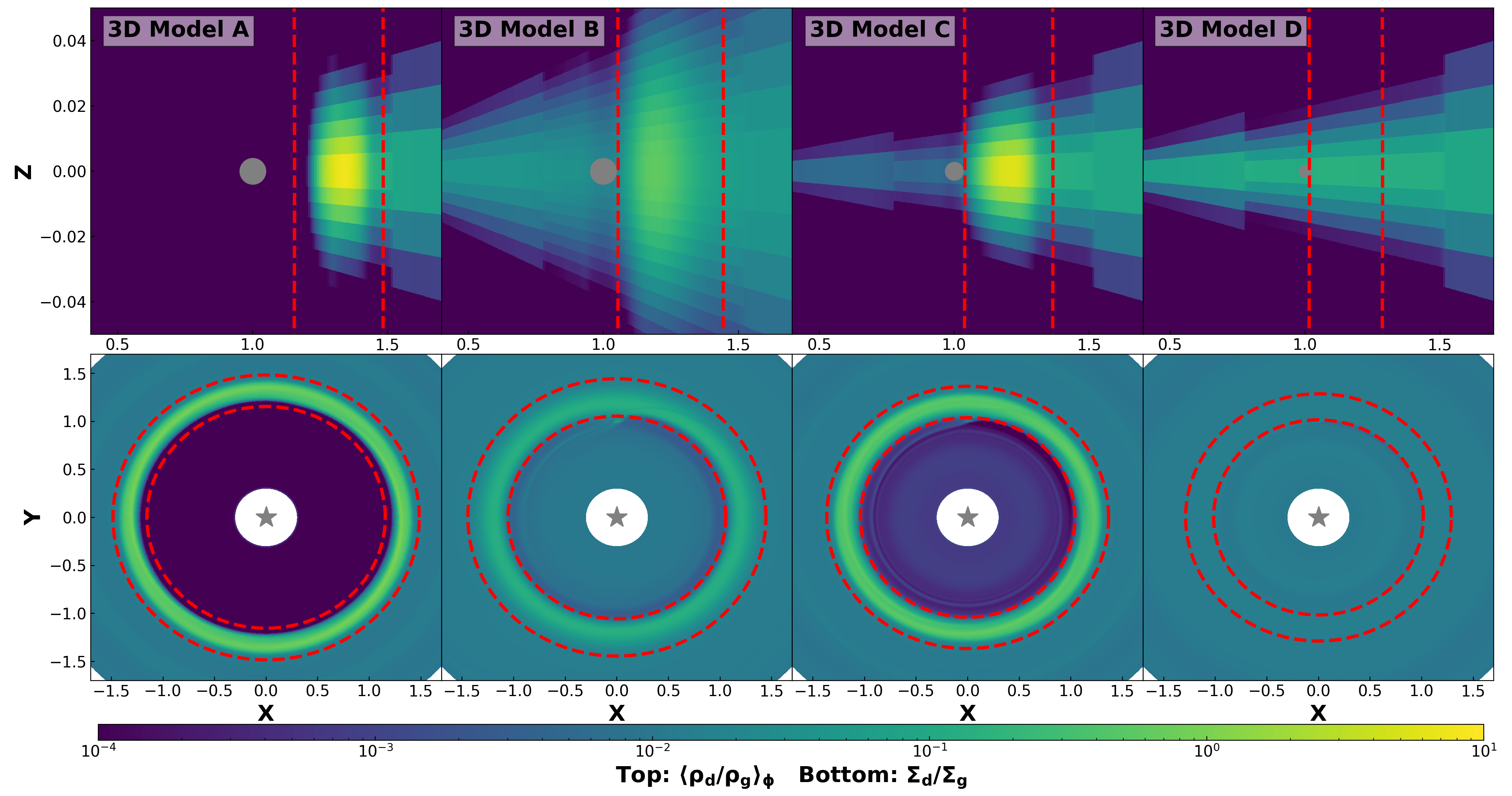}
\caption{
Similar to Figure~\ref{fig:3D_gas_densities}, but for the azimuthally averaged dust-gas density ratios $\langle \rho_\text{d}/\rho_\text{g} \rangle_\phi$ and dust-gas surface density ratios $ \Sigma_\text{d}/\Sigma_\text{g}$ for 3D models at 1500 orbits. Red dashed lines represent the inner boundary and the outer boundary of dusty rings.}
\label{fig:3D_dust_ratios}
\end{figure*}

\begin{figure*}
        \centering
        \includegraphics[width=\textwidth]{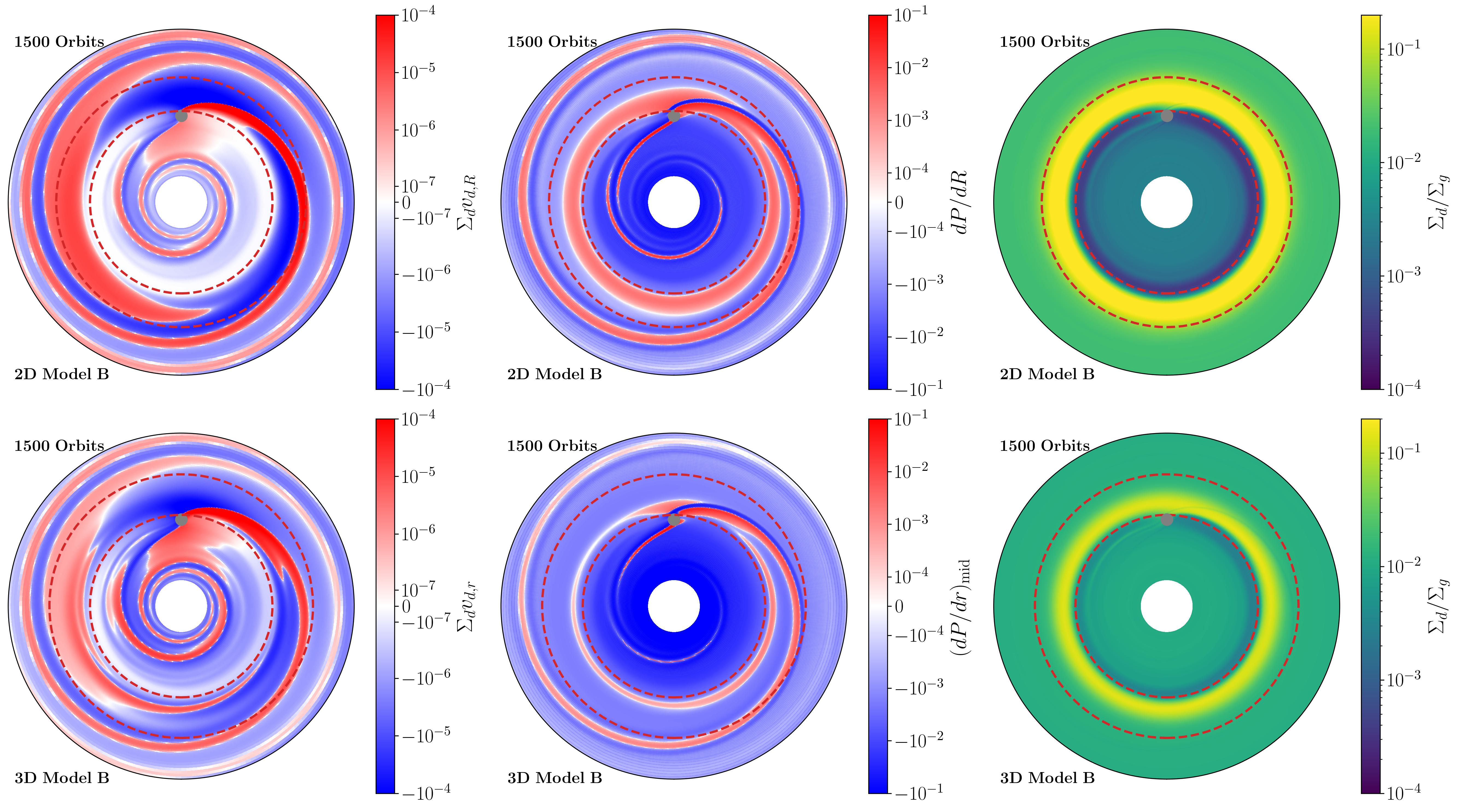}
        \caption{$r-\phi$ bird's eye view of radial dust flux $\Sigma_d v_{d,r}$ (left column), pressure radial gradient at the mid-plane $(dP/dr)_{\rm mid}$ (middle column) and dust-to-gas ratio $\Sigma_d/\Sigma_g$ (right column) for the 2D simulation (upper row) and 3D simulation (lower row) in Model B in Table~\ref{tab:Models}).
        Red dashed lines represent the inner boundary and the outer boundary we choose.
        Gray dots represents the location of the planet.}
        \label{figs: rphi_ModelA}
\end{figure*}

We also define the mass trapping ratio within the dust ring:

\begin{equation}
\epsilon_{\rm mass} (t)=\frac{\int_{t_\text{min}}^{t} \left(\dot{M}_{\text{ring,outer}}-\dot{M}_{\text{ring,inner}}\right)dt}{\int_{t_\text{min}}^{t} \dot{M}_{\text{ring,outer}}\,dt}.
\label{eq:massratio}
\end{equation}

Calculations of $\epsilon_{\rm flux}$ and $\epsilon_{\rm mass}$ 
both require a definition of the dust ring, 
which we choose from visual inspection. 
By definition, the dust trap is identified by locally high dust concentration. 
Determining the inner radius is therefore relatively straightforward and we simply 
select a position, by eye, where the dust density significantly drops close to the planet.
The location of the ring gradually shifts outward over time and so we define the inner edge of the ring after a few hundred orbits beyond which the ring remains relatively stationary.
This approach ensures that the chosen inner radius remains outside the actual boundary of the dust trap during its outward migration.
At the same time, we ensure that the inner edge is located slightly beyond one Hill radius $r_{\rm Hill}$ from the planet, so that all positions within the considered region remain at distances larger than the gravitational softening length $r_s$, thereby preserving the accuracy of the gravitational force. This setup also excludes the horseshoe region near the planet, effectively eliminating the influence of the Lagrangian trap on our calculations.
In contrast, selecting the outer radius is more challenging due to the continuous supply of dust from the outer region. We demonstrate in Appendix \ref{app:boundary} the robustness of our calculations of trapping efficiencies to the variation of $r_{\rm out}$.

\section{Results}~\label{sec:results}
In this section, we present the general morphology of gas and dust profiles for both 2D and 3D simulations in Section~\ref{subsec:morphology}, then calculate the dust trapping efficiency in Section~\ref{subsec:efficiency}.

\begin{figure*}[htp]
\centering
\includegraphics[scale=0.45]{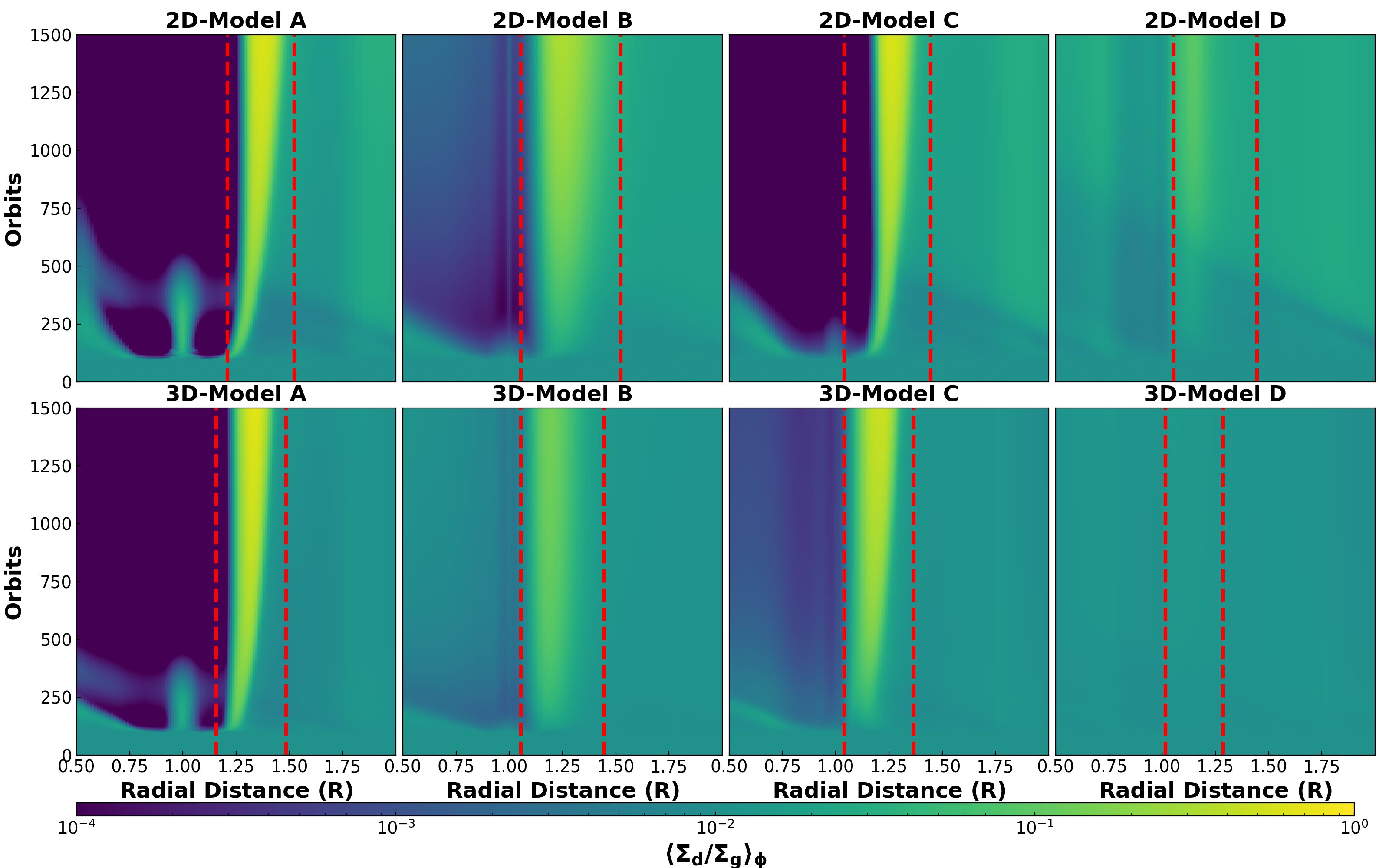}
\caption{The space-time ($R-t$) evolution of the azimuthally averaged dust-gas surface density ratios $\langle \Sigma_\text{d}/\Sigma_\text{g} \rangle_\phi$  for different 2D (top row) and 3D (bottom row) models. The x axes represent the radial distance while the y axes represent the orbit numbers counted at $R_\text{p} = 1$. Red dashed lines represent the inner boundary and the outer boundary of dusty rings for 2D and 3D cases.}
\label{fig:space_time_surface_density}
\end{figure*}

\begin{figure*}[htbp]
        \begin{minipage}[b]{0.5\linewidth}
            \centering
            \includegraphics[width=\columnwidth]{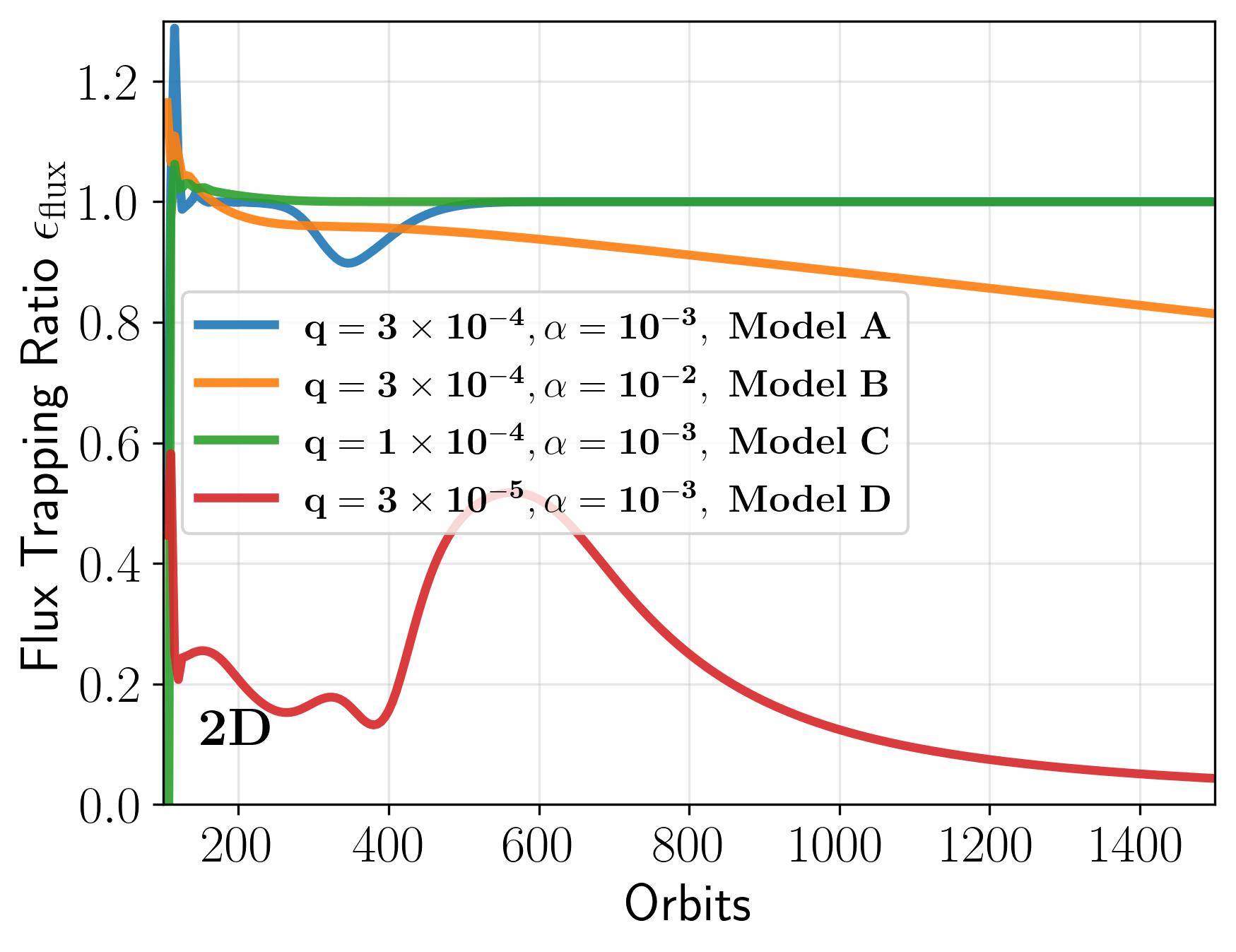}
        \end{minipage}
        \begin{minipage}[b]{0.5\linewidth} 
            \centering
            \includegraphics[width=\columnwidth]{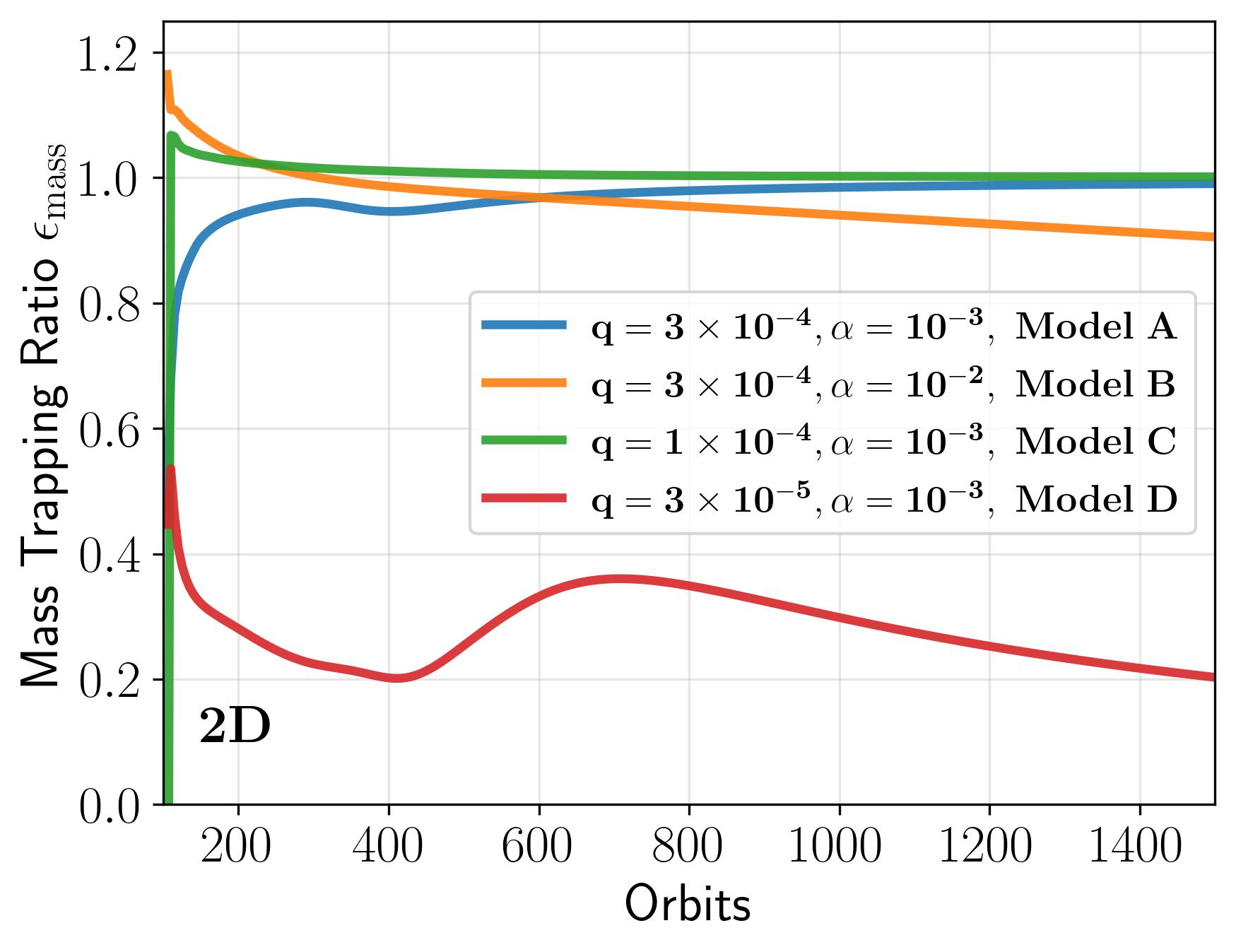}
        \end{minipage}
        \begin{minipage}[b]{0.5\linewidth}
            \centering
            \includegraphics[width=\columnwidth]{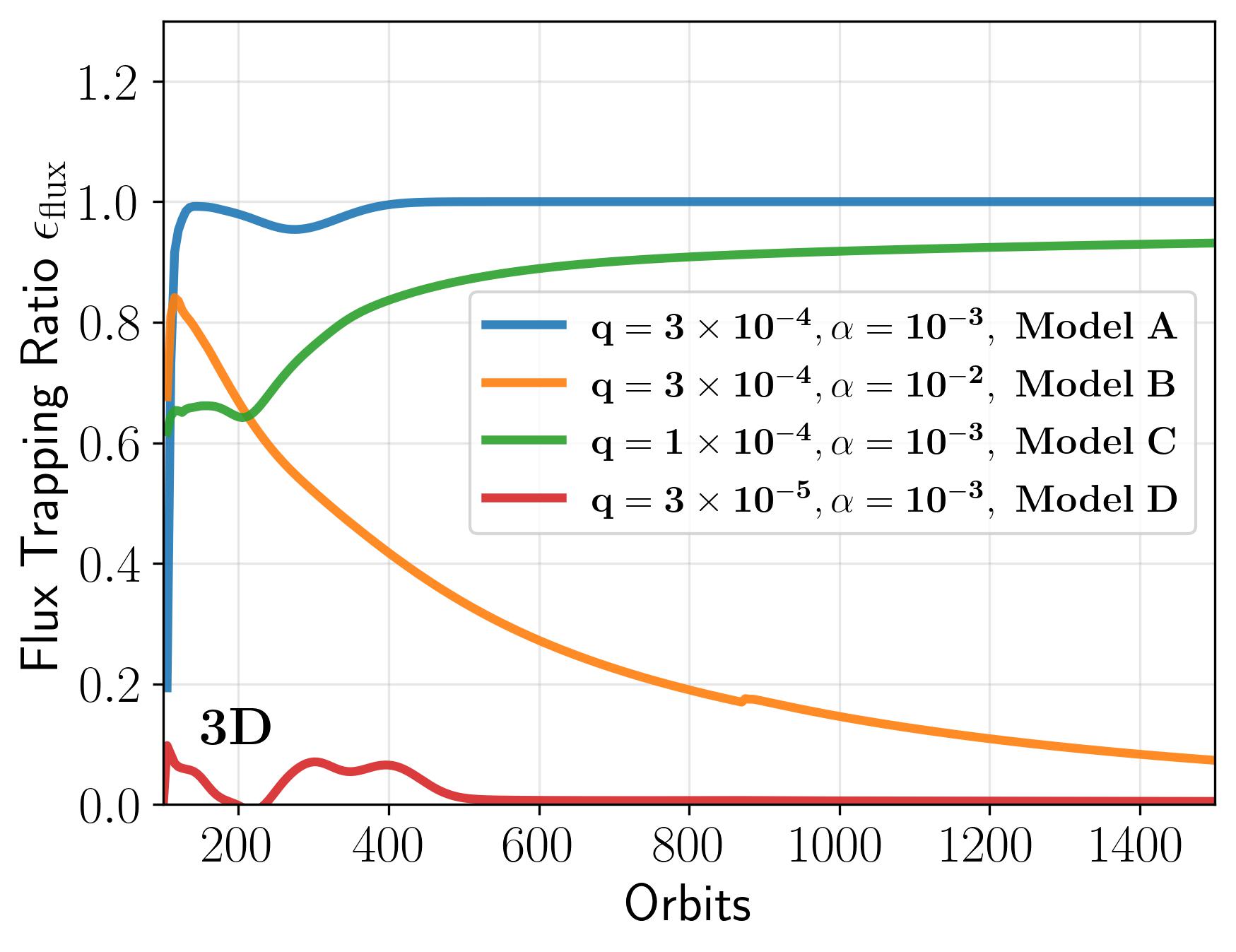}
        \end{minipage}
        \begin{minipage}[b]{0.5\linewidth} 
            \centering
            \includegraphics[width=\columnwidth]{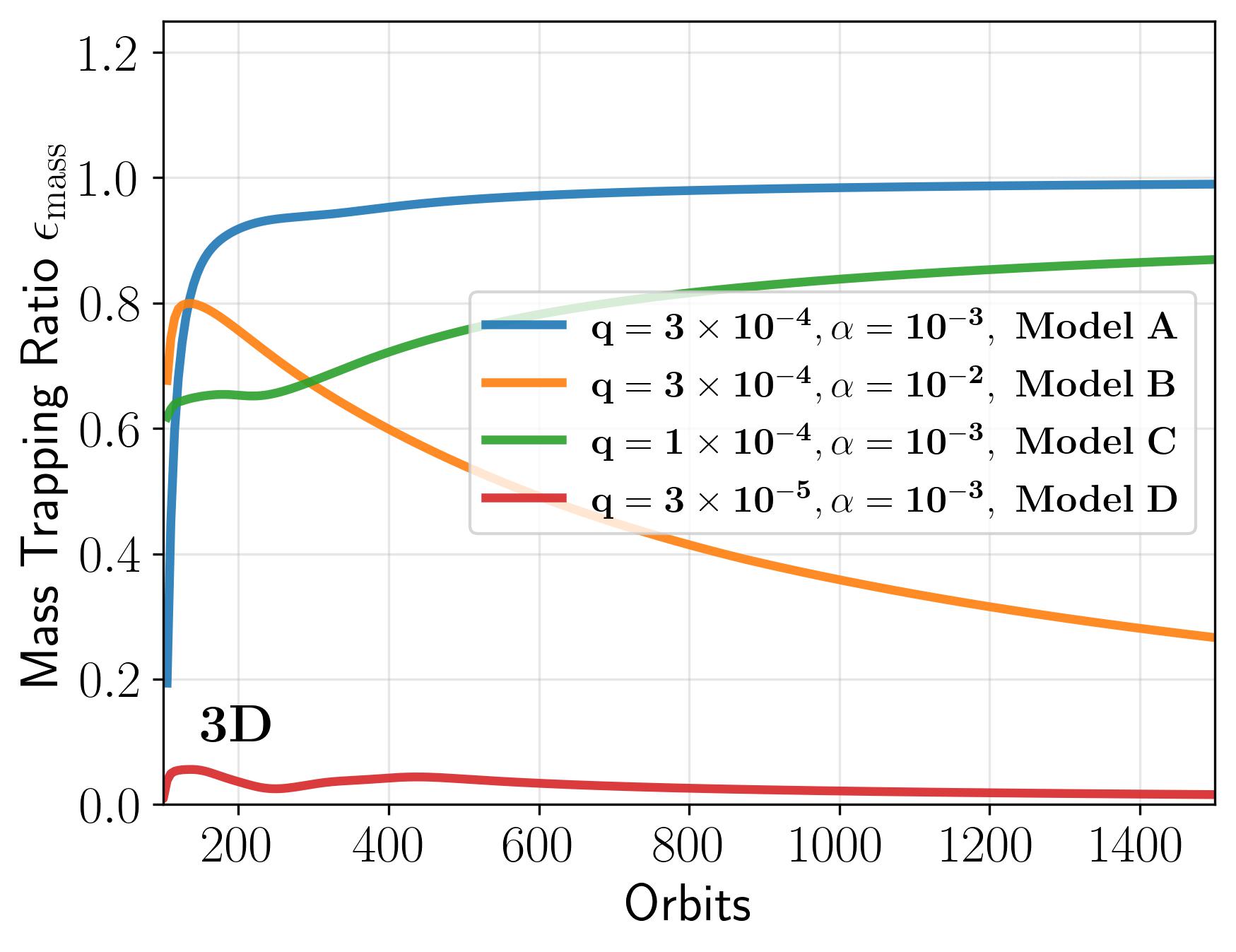}
        \end{minipage}
        \caption{
        Mass trapping ratio (left column) and flux trapping ratio (right column) defined in Eq.~(\ref{eq:trapratio}) and (\ref{eq:massratio}) as a function of time for the different 2D/3D simulation.}
        \label{figs:ratios}
\end{figure*}

\begin{figure*}[htbp]
        \centering
        \includegraphics[width=\textwidth]{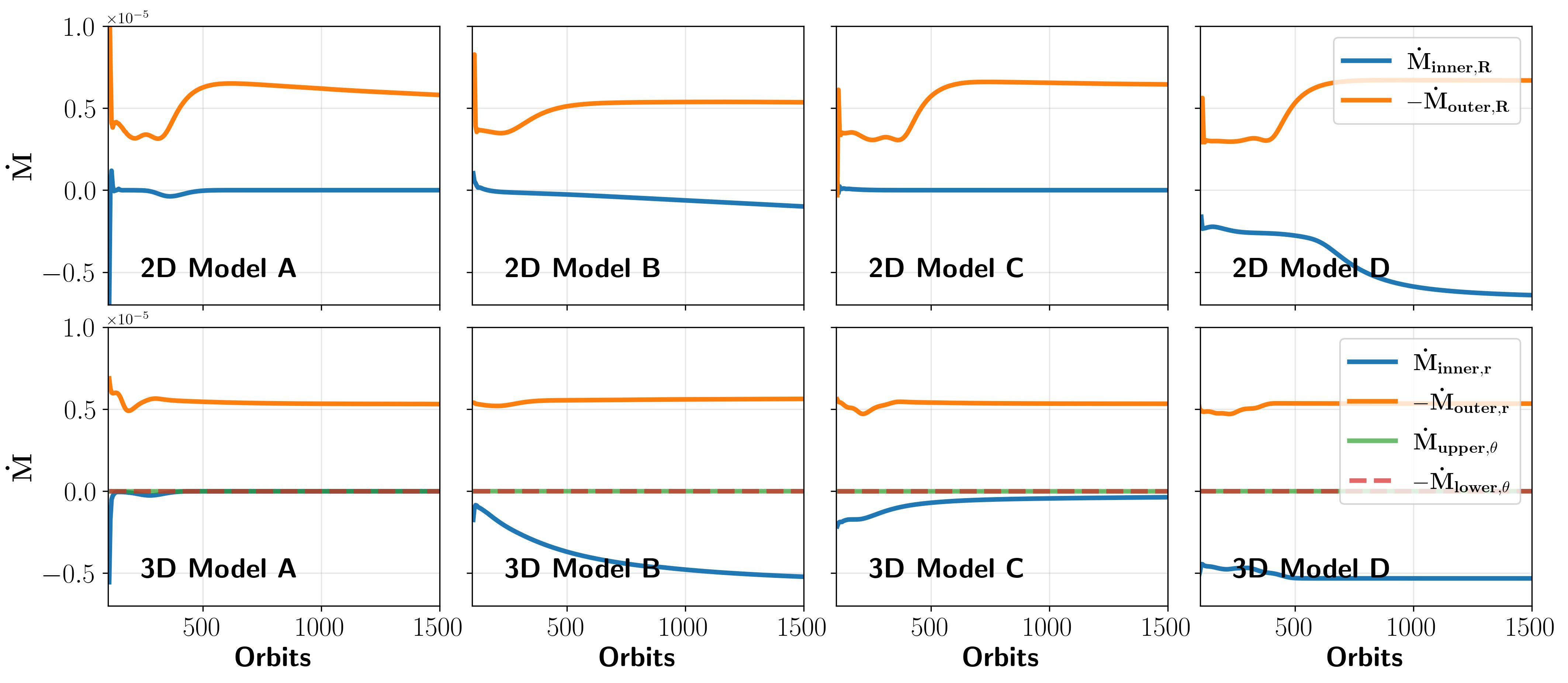}
        \caption{
        The mass flux $\dot{M}$ at the inner and outer boundaries of the dust trap is shown as a function of time for different 2D (upper row) and 3D (lower row) simulations. We plot $\dot{M}$ at the upper and lower boundaries as well to demonstrate that they are indeed zero at the edges of the computational domain.
        In the figure, positive values indicate flow into the dust trap.}
        \label{figs:flux}
\end{figure*}

\subsection{Gas and Dust Morphology}~\label{subsec:morphology}

Figure~\ref{fig:2D_density_ratios} presents the normalized gas density, $\Sigma_\text{g}/\Sigma_\text{g,init}$, and the dust-to-gas surface density ratio, $\Sigma_\text{d}/\Sigma_\text{g}$, for various 2D simulations after 1500 orbits. The simulations reveal that the embedded planet excites spiral density shocks~\citep{GoodmanRafikov2001}, which deposit angular momentum into the local disk material, leading to the formation of annular gaps.

The contrast of the spiral density waves decreases as the turbulent viscosity increases (Models A and B) or the planetary mass decreases (Models A, C, and D). Among all 2D models, the planet in Model A forms the deepest gaps, with $\Sigma_\text{g,gap}/\Sigma_\text{g,init} \sim 0.1$, consistent with \citet{Fung2014}, followed by the planets in Models C and B, where $\Sigma_\text{g,gap}/\Sigma_\text{g,init} \sim 0.5$. In contrast, the planet in Model D creates only a weak perturbation, with $\Sigma_\text{g,gap}/\Sigma_\text{g,init} \sim 0.9$. At the same time, secondary spirals are observed in the inner and outer gas disks~\citep{BaeZhu2018I,BaeZhu2018II}.

We construct the volumetric density profile such that the hydrodynamic properties of the 3D simulations match those of the 2D runs. To demonstrate this, we present azimuthally averaged surface density profiles in Figure~\ref{fig:Sigma_2D_3D}, excluding a small azimuthal region of width $\pm \Delta/R_p$, where $\Delta \equiv 2r_{\rm Hill}$, around the planet following \citet{FungChiang2016}. In both gap width and depth, the 3D results show good agreement with their 2D counterparts.

The pressure maxima at the gap edge can trap dust.
Based on Figure~\ref{fig:2D_density_ratios}, the dust-to-gas surface density ratios, $\Sigma_\text{d}/\Sigma_\text{g}$, in the inner disks of Models A and C drop below $10^{-4}$, indicating that dust is fully trapped in these two models. 
By contrast, we see more leakage of dust in Models B (high $\alpha$) and D (low planet mass).
Comparing 2D (Figure \ref{fig:2D_density_ratios})) with 3D (Figures \ref{fig:3D_gas_densities} and \ref{fig:3D_dust_ratios}), we see similar behavior of planet-opened gap and waves in the $r-\phi$ plane, 
although we see shallower gaps in the gas disk in 3D. As a result, we find heightened level of dust leakage for all our models except for Model A in 3D (see Figure \ref{fig:3D_dust_ratios}) compared to 2D. Vertically, Figures 2 and 3 show how the dust disk is much more concentrated toward the midplane as expected from settling. Quantitatively, we find the dust disk scale height reaches $\sim$0.1 times the gas disk scale height, consistent with theoretical expectations \citep[e.g.,][]{Fromang09,Chiang10}.

The upper row of Figure \ref{figs: rphi_ModelA} illustrates the birds-eye view of the dust radial flux $\Sigma_d v_{d,R}$, gas pressure gradient at the midplane $(dP/dR)_\text{midplane}$, and dust-to-gas surface denisty ratio $\Sigma_d/\Sigma_g$ for the fiducial model (Model A). As expected for planetary perturbation, both the dust flux and the pressure gradient show non-axisymmetric and spiral features, in addition to a strong positive to negative $dP/dR$ just outside the planet's orbit, corresponding to the clear pressure maximum with enhanced $\Sigma_d/\Sigma_g$. Similar non-axisymmetric dust flux patterns were observed in the 3D simulations by \citet{binkert23}.
The radial flux shows clearly how the dust leaks from the planetary gap edge through the density waves. In general, we see roughly an order of magnitude enhancement in the concentration of the dust that leaked through the pressure trap (and stronger radial flux) in 3D compared to 2D at all times,
as illustrated in the space-time plot of dust-to-gas surface density ratio in Figure \ref{fig:space_time_surface_density}, with the exception of Model A (lowest $\alpha$ and highest planet mass in our study) which features a perfect trap in both 2D and 3D. This is consistent with \citet{bi21, bi23}, who highlighted the role of meridional flows in enhancing vertical and radial dust transport.
We note the short-lived concentration of dust at the radial location of the embedded planet in Model A that facilitate a short burst of dust advection from the exterior to the interior of the planet's orbit. This concentration is a transient Lagrangian trap which we discuss in more detail in Section \ref{subsec:lagrangian}.\footnote{Our simulation produces only one dust ring instead of multiple rings that have been reported in previous studies \citep[e.g.,][]{DongLi2017,DongLi2018} because our adopted $\alpha > 10^{-4}$. A lower $\alpha$ is required to allow the shock of secondary and tertiary spiral density waves to be properly captured for the creation of multiple rings generated by a single embedded planet.}

\subsection{Trapping Efficiency}~\label{subsec:efficiency}

Based on our definition of trapping efficiency, a perfect trap (i.e., $\dot{M}_{\rm ring, inner} = 0$ at all times) corresponds to $\epsilon_{\rm mass} = \epsilon_{\rm flux} = 1$. 
Since our initial condition already starts with some amount of dust distributed uniformly within the disk, $\dot{M}_{\rm ring, inner}$ can be positive (i.e., outward flux) initially leading to both $\epsilon$ being greater than unity. If and once the dust flux across the pressure trap reaches a steady state (i.e., the pressure maximum has trapped as much dust as it can trap), we expect $\epsilon_{\rm flux} = 0$ but $\epsilon_{\rm mass} \neq 0$ albeit decreasing with time as the latter metric is cumulative. If the $\epsilon$'s remain $<1$ at all times, it implies a given dust trap is leaky.

Figure \ref{figs:ratios} demonstrates how the $\epsilon$'s change with time over the range of $M_p$, $\alpha$ and 2D vs.~3D.
In 2D, we observe nearly perfect dust trapping in all our models with $q \equiv M_{\rm p}/M_\star \geq 10^{-4}$. We observe an initial phase of $\epsilon > 1$ from the existing dust at the inner edge of the ring collecting toward the center of the pressure bump (see $\dot{M}_{\rm inner,R} > 0$ for 2D Models A--C in Figure \ref{figs:flux}). While Models A \& C feature a nearly constant $\epsilon \sim 1$, Model B features a decline in $\epsilon$'s with time. The sudden drop in $\epsilon_{\rm flux}$ in Model A at 300--400 orbits is from the short-lived collection of dust at the radial location of the planet, sourced from the trap; see also Figure \ref{fig:space_time_surface_density}. This is a transient Lagrangian trap that will be discussed further in Section \ref{subsec:lagrangian}. Given the increase in $\Sigma_d/\Sigma_g$ with time within the dust trap apparent in Figure \ref{fig:space_time_surface_density}, we do not think Models A--C have reached a steady-state dust trap yet.
Figure \ref{figs:flux} shows $\dot{M}_{\rm inner, R}$ is becoming increasingly negative and we expect that if evolved long enough, 2D Model B will reach a dust trap steady state and $\epsilon_{\rm flux} = 0$. 
By contrast, Model D maintains low $\epsilon$'s at all times suggesting highly inefficient trap, consistent with the relatively low $\Sigma_d/\Sigma_g$ within the trap and shallow gap shown in Figure \ref{fig:space_time_surface_density}.

Among 2D simulations, we see a reduction in both $\epsilon_{\rm mass}$ and $\epsilon_{\rm flux}$ with time with higher $\alpha$ (more turbulent diffusion) and with lower planet mass (shallower planetary gap and therefore weaker pressure maximum in gas). 
The aforementioned trend of trapping efficiency with respect to $q$ and $\alpha$ stay the same when we observe 3D simulations, albeit the magnitude of $\epsilon_{\rm mass}$ and $\epsilon_{\rm trap}$ being even more reduced compared to 2D. Even in 3D, Model A achieves a perfect trap.
For all other models, $\epsilon$'s are significantly more reduced in 3D compared to 2D, consistent with more filled in $\Sigma_d/\Sigma_g$ inside the gap seen in Figure \ref{fig:space_time_surface_density} for Models B, C, and D. Among these 3 models, Model C achieves $\epsilon$'s that gradually increase to near unity.

In order to better understand the non-smooth time-variability of the trapping efficiency showcased in Figure \ref{figs:ratios}, we look at how the mass fluxes $\dot{M}$ across the boundaries of the dust trap evolve with time (see Figure \ref{figs:flux}).
In both 2D and 3D simulations, there exists an initial transient state from the moment the dust is released where the dust flux entering the dust trap does not yet stabilize ($\sim$300 orbits for 3D and $\sim$400--500 orbits for 2D).
This initial transiency results in fluctuations and peaks in the trapping efficiency curves for different trapping efficiencies.
For Model D, the dust trap is very leaky, causing the mass flux across the inner and outer boundaries to be comparable, enhancing the level of fluctuations seen in $\epsilon_{\rm flux}$ vs.~time.

\begin{figure*}[htbp]
        \centering
        \includegraphics[width=\textwidth]{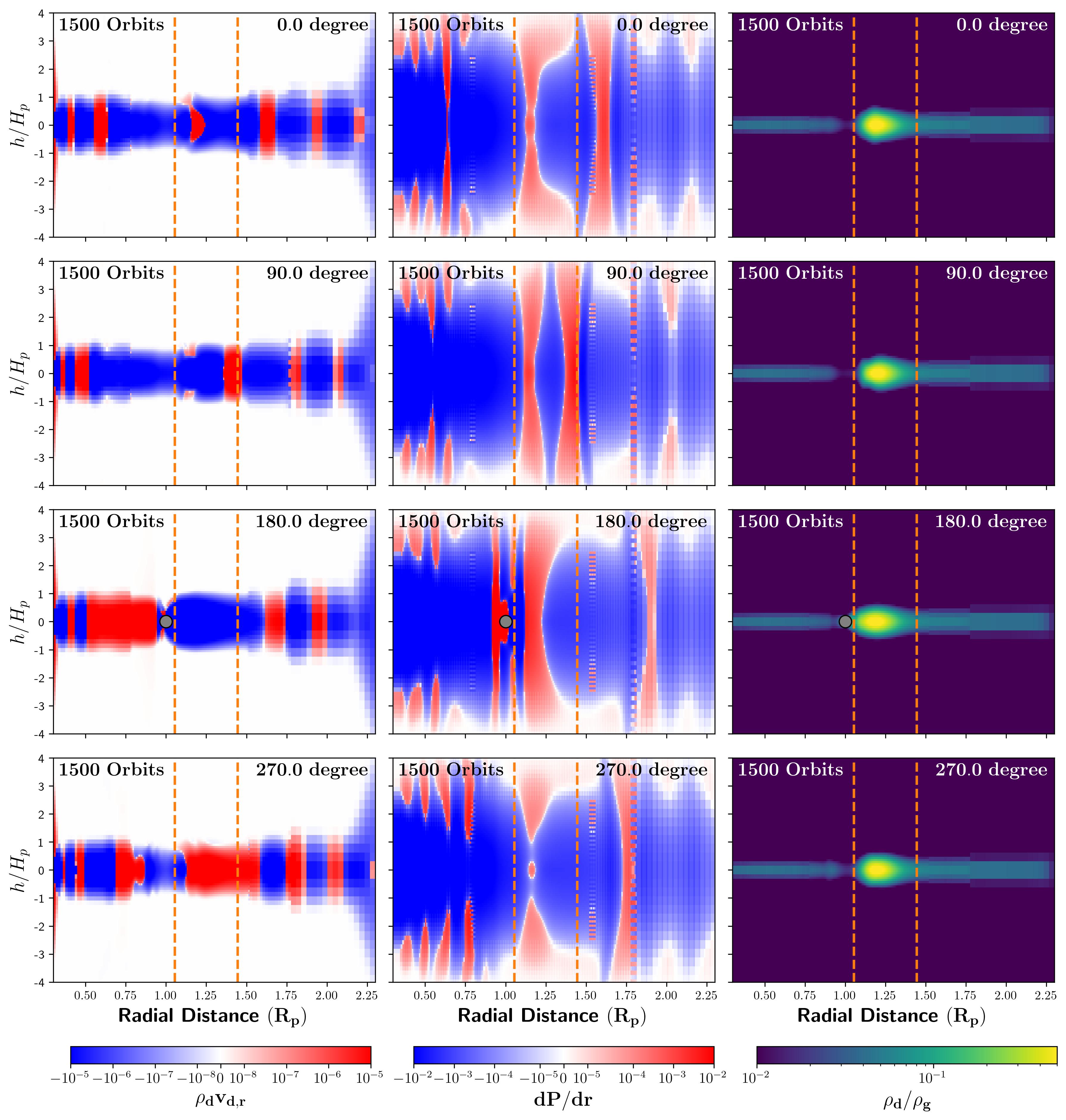}
        \caption{
        $r-\theta$ slice of radial dust flux $\rho_d v_{d,r}$ (left column), radial pressure gradient $dP/dr$ (middle column) and dust-to-gas ratio $\rho_d/\rho_g$ (right column) for the Model B in Table~\ref{tab:Models} at different azimuthal angles (different rows, note that the planet sits at $\phi=180^\circ$, gray dots represent the location of the planet.). When calculating the pressure radial gradient $dP/dr$, we use an interval of five grids to prevent drastic variations due to numerical resolution.
        Discontinuities in the middle and right columns appear at the boundaries of refinement.
        Orange dashed lines represent the inner and the outer boundaries of the dust trap.
        All quantities shown here are in normalized units; $H_p$ denotes the gas scale height at the planet's location.}
        \label{figs:3Drtheta_fiducial}
\end{figure*}

\begin{figure*}[htbp]
        \begin{minipage}[c]{0.4\linewidth} 
            \centering
            \includegraphics[width=\columnwidth]{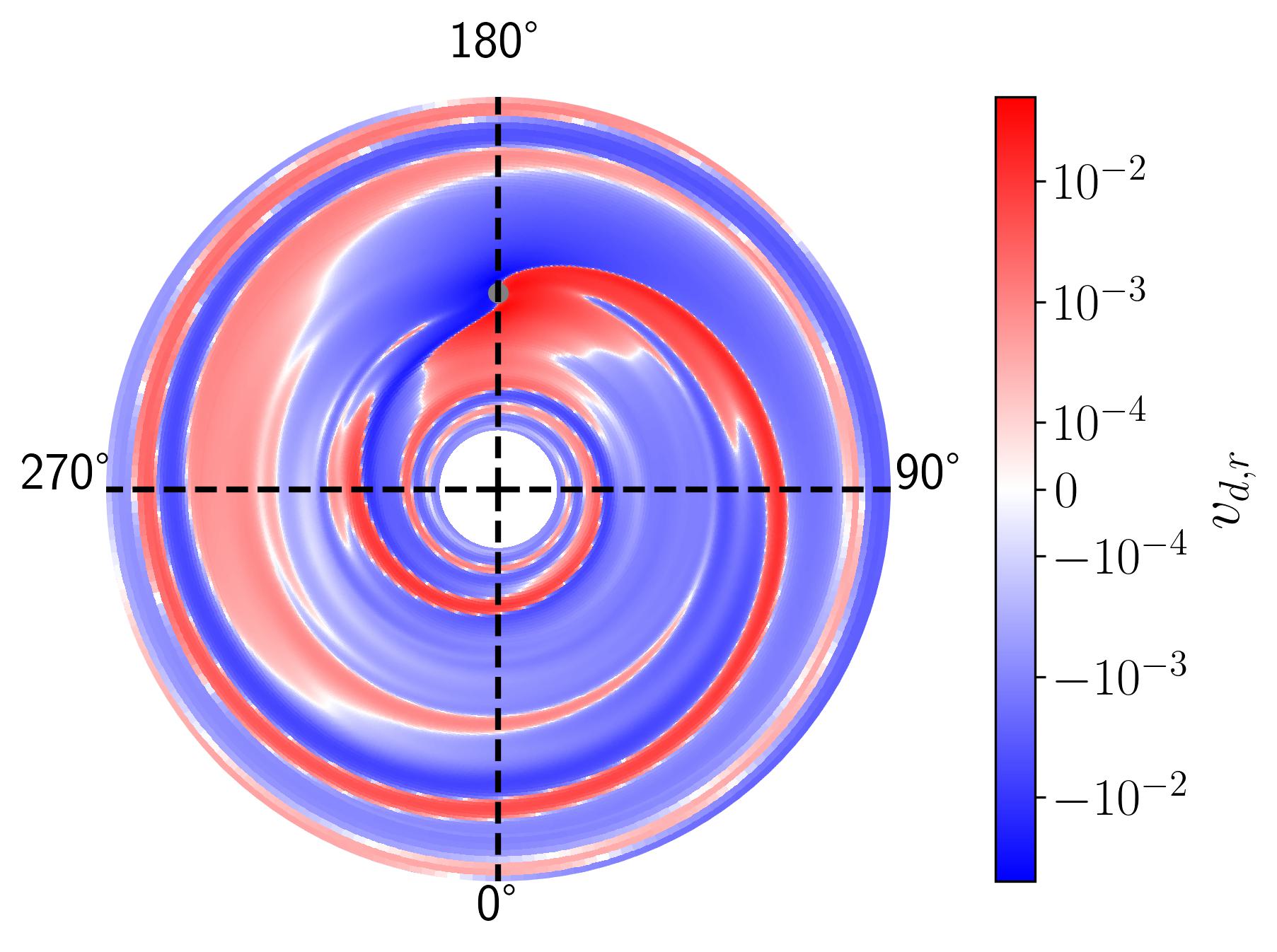}
        \end{minipage}
        \begin{minipage}[c]{0.6\linewidth}
            \centering
            \includegraphics[width=\columnwidth]{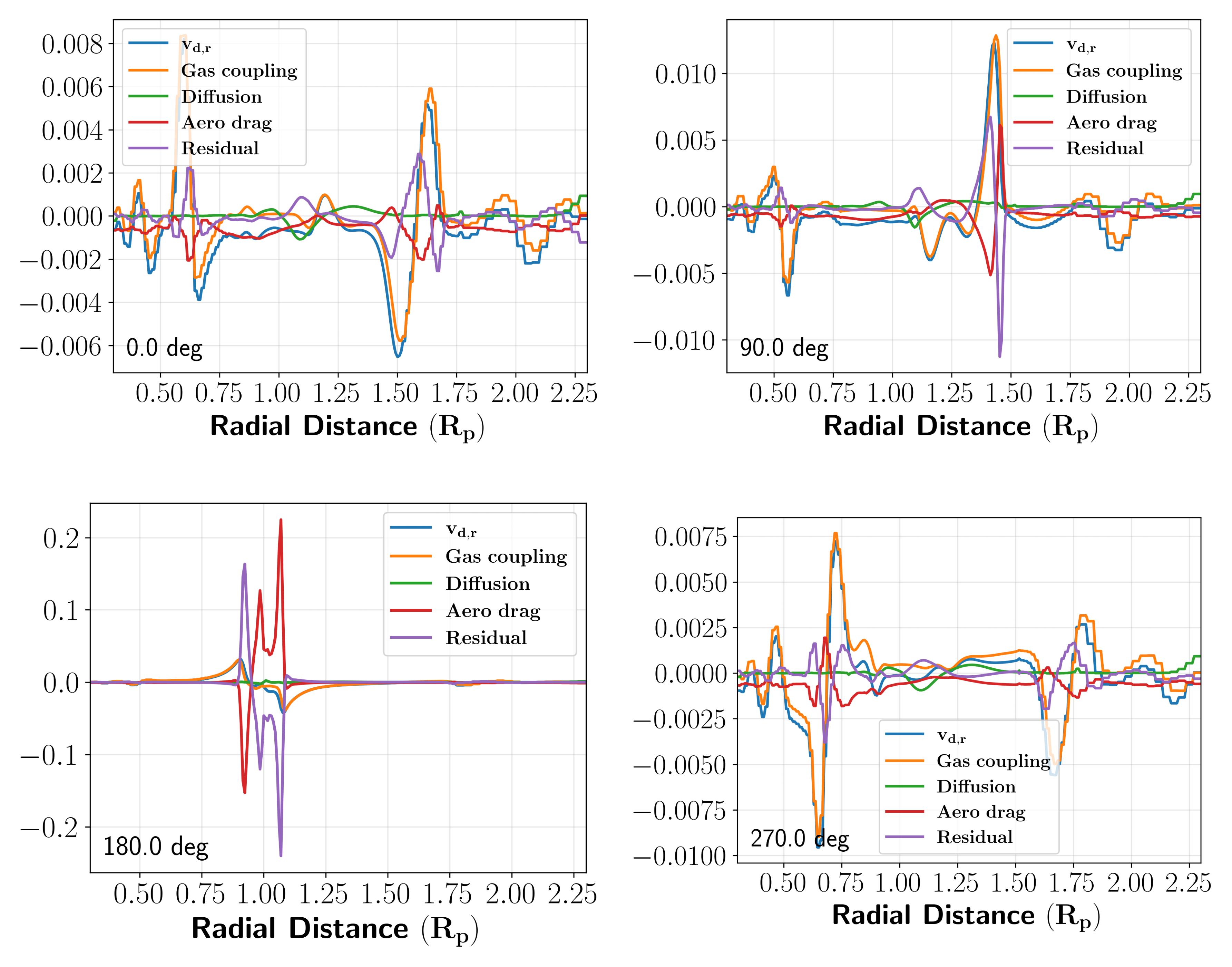}
        \end{minipage}

        \caption{
        Decomposition of dust radial motion at different azimuthal angles in Model B. 
        The bird-eye view figures on the left side of each panel show the dust radial velocity in the midplane, where our decomposition is most effective. 
        The black dashed lines in the figures represent the selected azimuthal slices, ranging from 0$^\circ$ to 270$^\circ$ from top left to bottom right. 
        Note that we define the planet's opposite direction as 0$^\circ$, with the counterclockwise direction being positive.
        The plots on the right of each panel shows the velocity decomposition along the azimuthal slice indicated by the black dashed line on the left, plotted as a function of radial position $r$. 
        The blue solid line represents the actual radial velocity $v_{\rm d,r}$ of the dust (evaluated along the dashed lines at corresponding azimuthal angle shown in the leftmost panel). The other solid lines represent the diffusion radial velocity $v_{\rm diff}$ obtained directly from the simulation and those defined by Eqs.~(\ref{eq:vcouple}-\ref{eq:vresidual}) as $v_{\rm coupl}$, $v_{\rm drag}$, and $v_{\rm residual}$.}
        \label{figs:Decomposition_Model_A}
\end{figure*}

\begin{figure*}[htbp]
        \begin{minipage}[c]{0.4\linewidth} 
            \centering
            \includegraphics[width=\columnwidth]{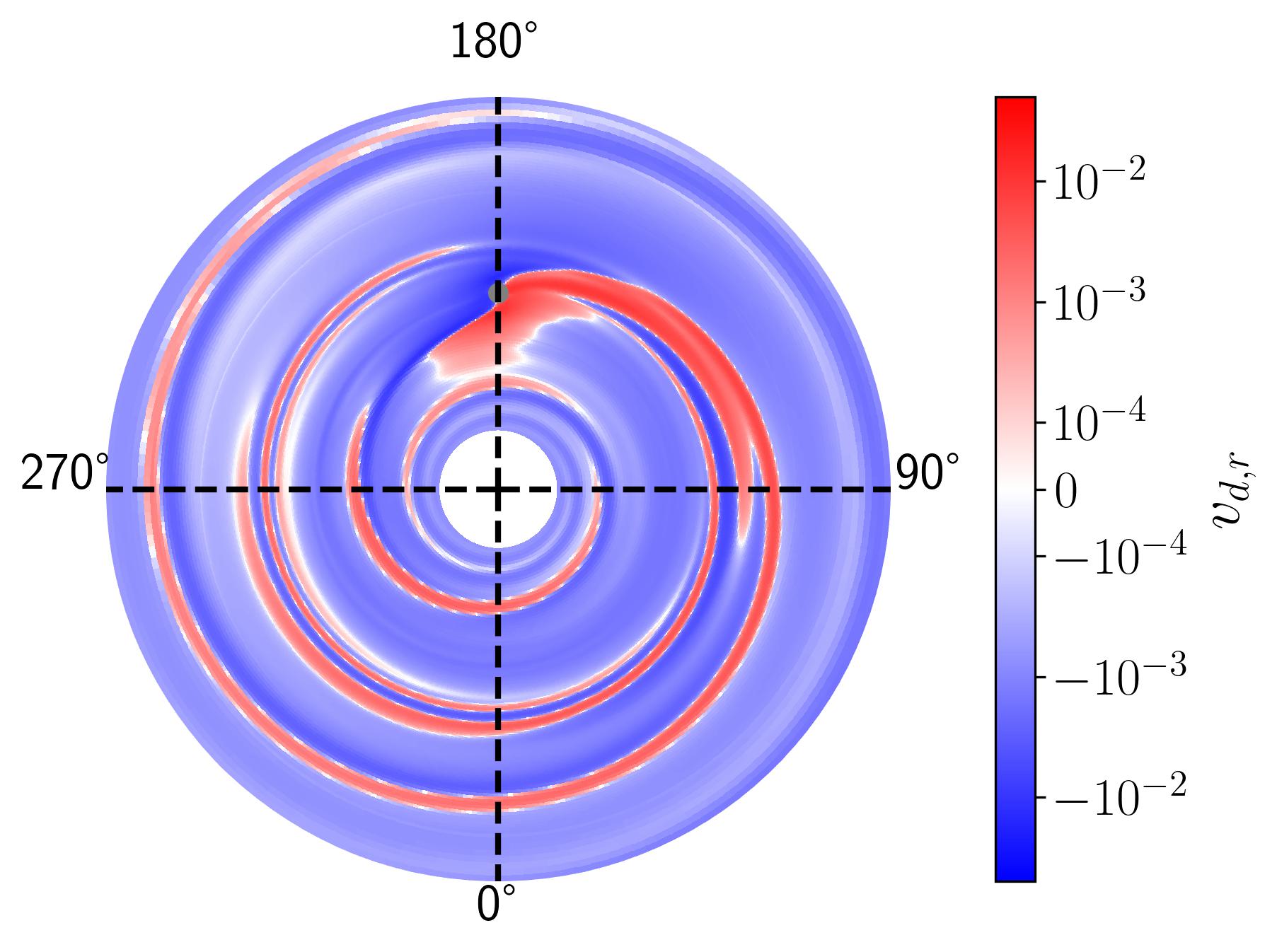}
        \end{minipage}
        \begin{minipage}[c]{0.6\linewidth}
            \centering
            \includegraphics[width=\columnwidth]{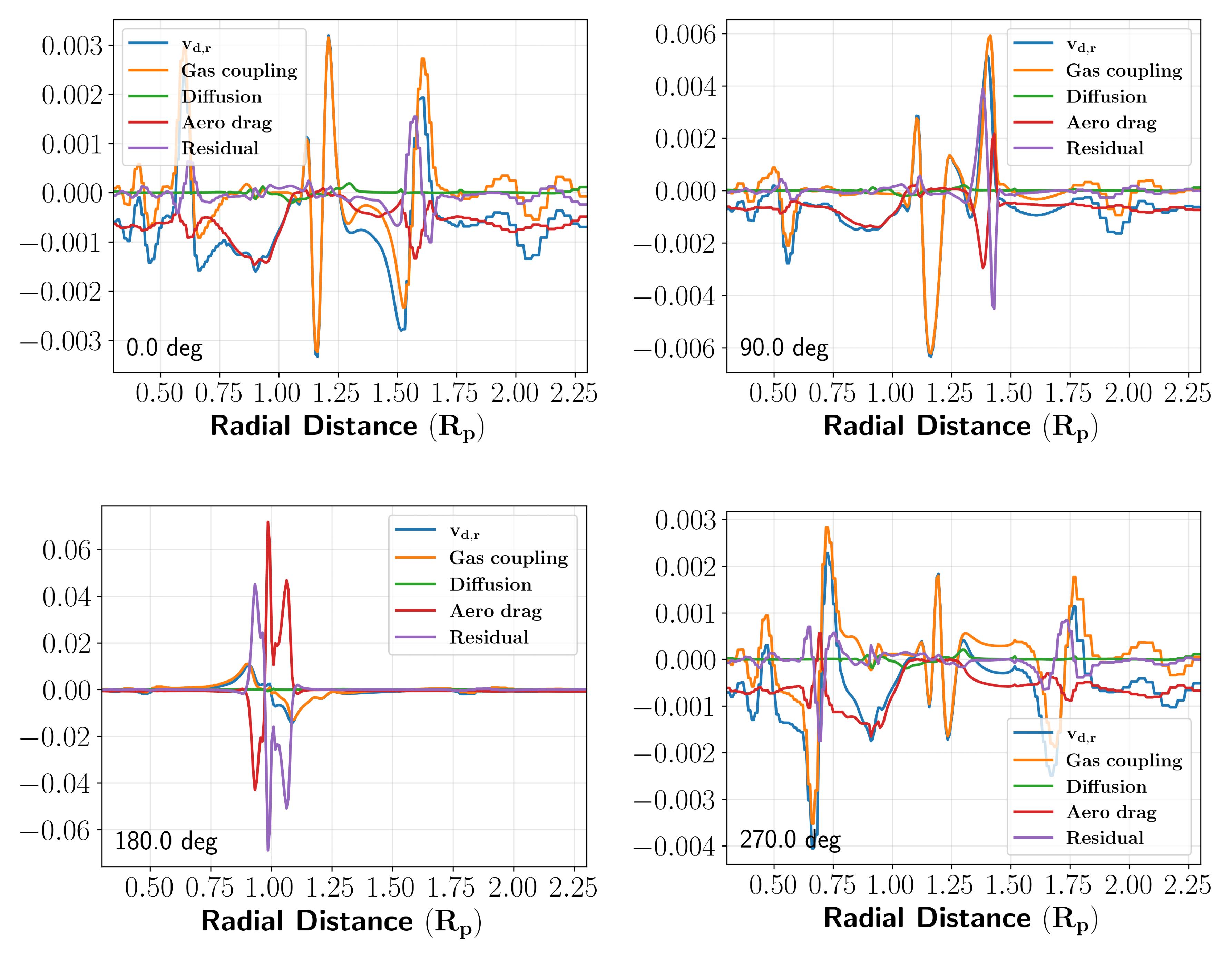}
        \end{minipage}
    \caption{
    Same as Figure~\ref{figs:Decomposition_Model_A}, except for Model C.}
        \label{figs:Decomposition_Model_C}
\end{figure*}

\section{Discussion}
\label{sec:discussion}

Our numerical results find that the dust trap becomes more leaky at higher levels of viscous diffusion and when the planet creating the trap has a sufficiently low mass. This is because, in these cases, the planetary gap is shallower.  At higher viscous $\alpha$, the perturbation in the gas is more quickly erased, producing a shallower planetary gap and therefore a weaker pressure trap. Similarly, a lower-mass planet induces weaker perturbations in the disk, also leading to a shallower gap.

While this trend of increasingly leaky dust trap with respect to increasing $\alpha$ and decreasing planet mass holds in both 2D and 3D, 3D simulations generally find the trap to be more leaky compared to 2D. To understand why, we analyze the dust and the gas motion in the $r-\theta$ plane, as illustrated in Figure \ref{figs:3Drtheta_fiducial}. We observe complex non-axisymmetric behavior with the degrees of radial influx of dust grains into the inner disk varying wildly at different azimuthal angles. At 0$^\circ$ (see Figure \ref{figs:Decomposition_Model_A} for the definition of azimuthal angles), we see a pocket of a dust trap exterior to the planet's orbit close to the midplane which disappears completely as we sweep through 90$^\circ$ to 180$^\circ$ transitioning to bulk radial inward motion. At 270$^\circ$, we find a large region of outward dust motion. 
Curiously, the radial motion of grains does not correlate with the radial pressure gradient map depicted in the middle column of Figure \ref{figs:3Drtheta_fiducial}, suggesting that aerodynamic drag may not be the dominant factor that determines the dynamics of dust in the presence of planetary perturbation. We discuss in the next section the dominant cause of the complex dust motion. 

\subsection{Decomposition of dust radial velocity}
\label{sec:decomp}

To identify the leading order term that drives the radial motion of dust grains, we decompose the radial velocity of dust into its coupling to gas flow $v_{\rm coupl}$, aerodynamic drag $v_{\rm drag}$, and diffusion $v_{\rm diff}$, the latter of which is computed directly from Athena++ following our fluid equations (see Section \ref{sec:method}). Direct perturbation of dust grains by the embedded planet will also play a role which we subsume under residual velocity $v_{\rm residual}$ for simplicity. Under steady-state condition, we define these components as follows:

\begin{equation}
    v_{\rm coupl, r}= \frac{v_{\rm g, r}}{1+St^2};
    \label{eq:vcouple}
\end{equation}

\begin{equation}
    v_{\rm drag, r}= -\frac{2(v_{\rm d, \phi}-v_{\rm g, \phi})}{St};
    \label{eq:vaero}
\end{equation}

\begin{equation}
    v_{\rm residual, r}= v_{\rm d,r}-v_{\rm coupl, r}-v_{\rm diff, r}-v_{\rm drag, r},
    \label{eq:vresidual}
\end{equation}
where $v_{\rm g, r}$, $v_{\rm g, \phi}$, and $v_{\rm d,r}$ are the numerically computed gas radial, gas azimuthal, and dust azimuthal velocity, respectively. Figure \ref{figs:Decomposition_Model_A} illustrates the complex behavior in our Model B. 
Focusing at $r \sim 1.2$ where we see a dust trap, the radial dust motion appears to be set by the gas coupling at all azimuthal angles. 
This strong dust-gas coupling driving dust redistribution aligns with the findings of \citet{binkert21}.
At 180$^\circ$, we see a huge spike in aerodynamic drag component which is completely negated by the ``residual'' motion. This is likely reflecting the dust grains aerodynamically reacting to the strong waves which in turn are generated by planetary perturbation. Similarly for Model C, we observe that dust coupling to gas motion remains the leading order term setting the radial motion of dust grains. The fact that the dust motion is dominated by its coupling to gas motion can also be discerned visually by the fact that the inward-outward alternating flux in the r-$\phi$ space closely follows the spiral pattern generated in the gas disk by the perturbing planet.

We further note that in 3D simulations, the dust trap features a complex morphology in the vertical direction.
The traps appear as pockets, sometimes close to the midplane (0$^\circ$ azimuthal angle) and sometimes off the plane (90$^\circ$). In other words, dust motion is both azimuthally and polodially non-uniform. This complexity in the vertical geometry of the dust trap is not captured in 2D simulations giving rise to dust traps being more leaky in 3D simulations compared to 2D simulations.

\subsection{Lagrangian traps}
\label{subsec:lagrangian}

Our study primarily focuses on smaller mass planets, as massive planets often form nearly perfect dust traps. 
However, our simulation results reveal a strong correlation between the occurrence of a Lagrangian trap and the effectiveness of the pressure bump as a dust trap. 
Specifically, when the pressure bump acts as a perfect trap, the Lagrangian trap typically forms. 
Conversely, if there is any leakage in the pressure bump, the Lagrangian trap does not form, allowing dust to pass through the planet and disperse into the inner region. 

\begin{figure}[htbp]
        \begin{minipage}[b]{\linewidth} 
            \centering
            \includegraphics[width=\columnwidth]{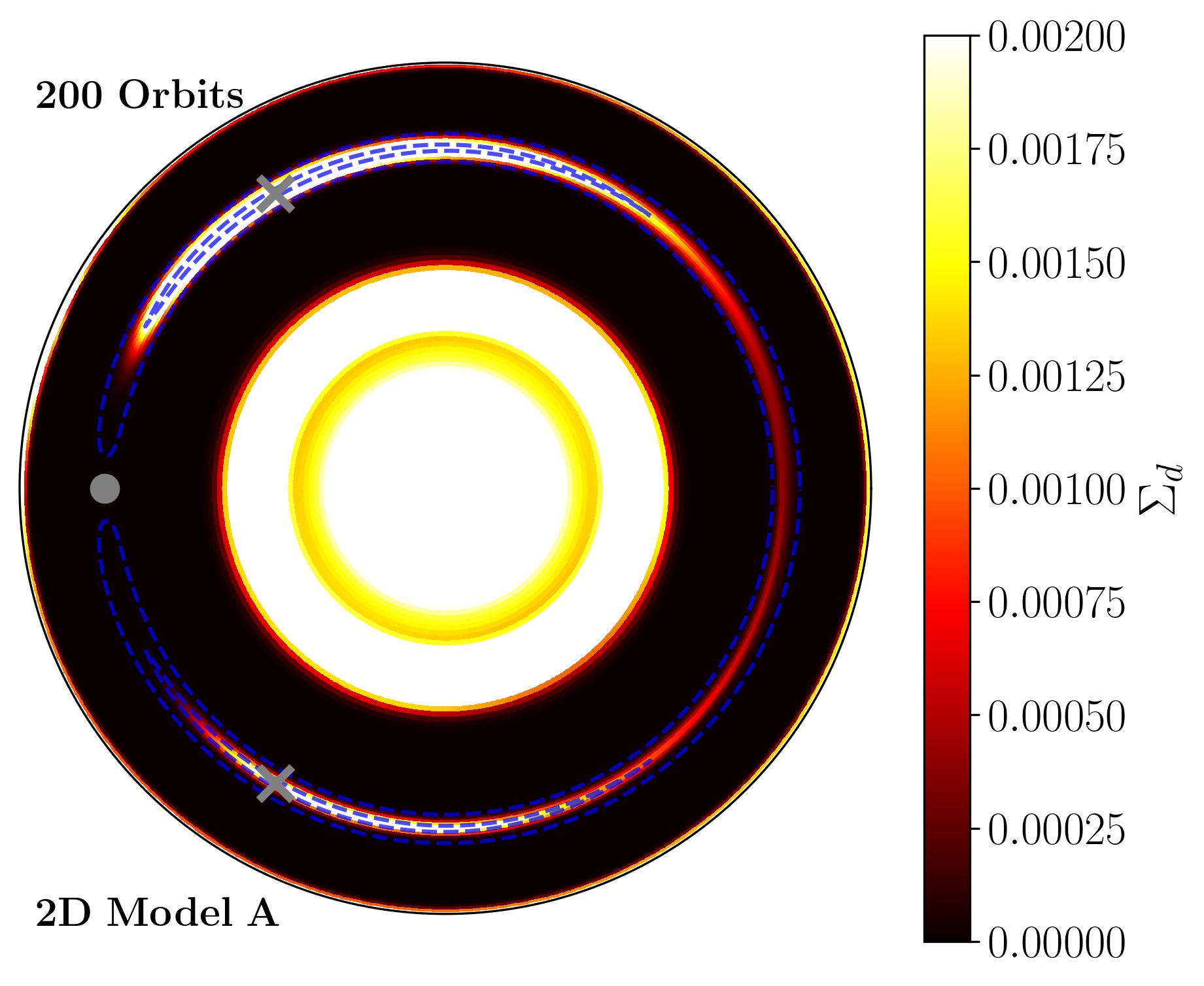}
        \end{minipage}
        \begin{minipage}[b]{\linewidth}
            \centering
            \includegraphics[width=\columnwidth]{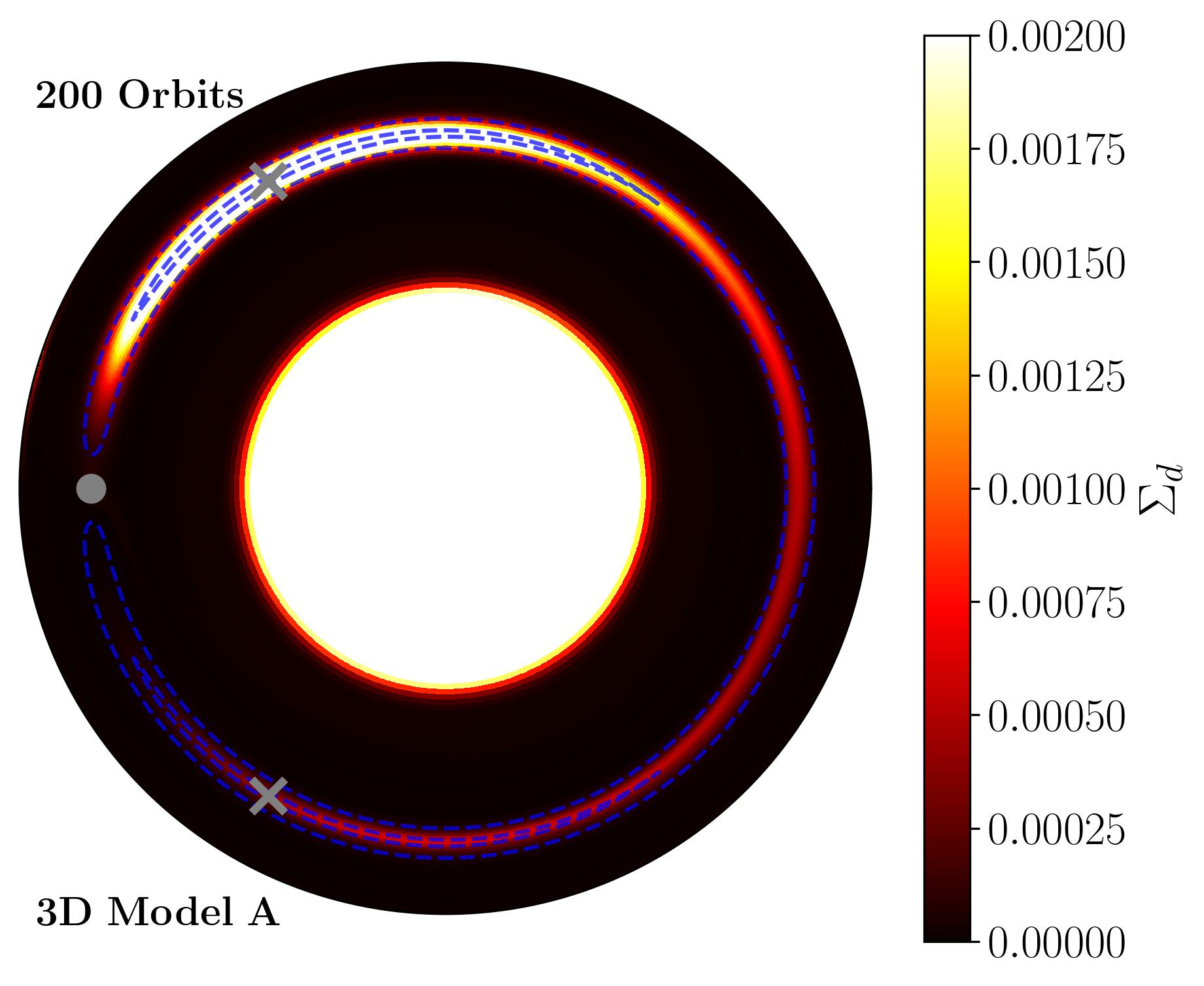}
        \end{minipage}
        \caption{Dust surface density of Lagrangian trap at 200 orbits both in 2D simulation (upper panel) and 3D simulation (lower panel) in Model A.
        The blue contour is an equipotential line, while the gray `x' marks indicate the locations of L4 and L5.}.
        \label{figs:Lag_Model_A}
\end{figure}

Figure~\ref{figs:Lag_Model_A} illustrates the distribution of dust surface density around the planet’s orbit after 200 orbits, highlighting Lagrangian dust traps produced by a subthermal mass planet previously seen in simulations as well \citep[e.g., see Fig. 11 in][]{DongLi2018}.
The 3D results exhibit a more pronounced asymmetry compared to the 2D results. 
Specifically, the surface density at L4 is lower than at L5 and tends to be more broadly distributed along the horseshoe track, rather than being concentrated at L4 and L5.
Additionally, the 3D Lagrangian trap appears more short-lived than its 2D counterpart (Figure \ref{fig:space_time_surface_density}).

\begin{figure}[htbp]
        \centering
        \includegraphics[width=\columnwidth]{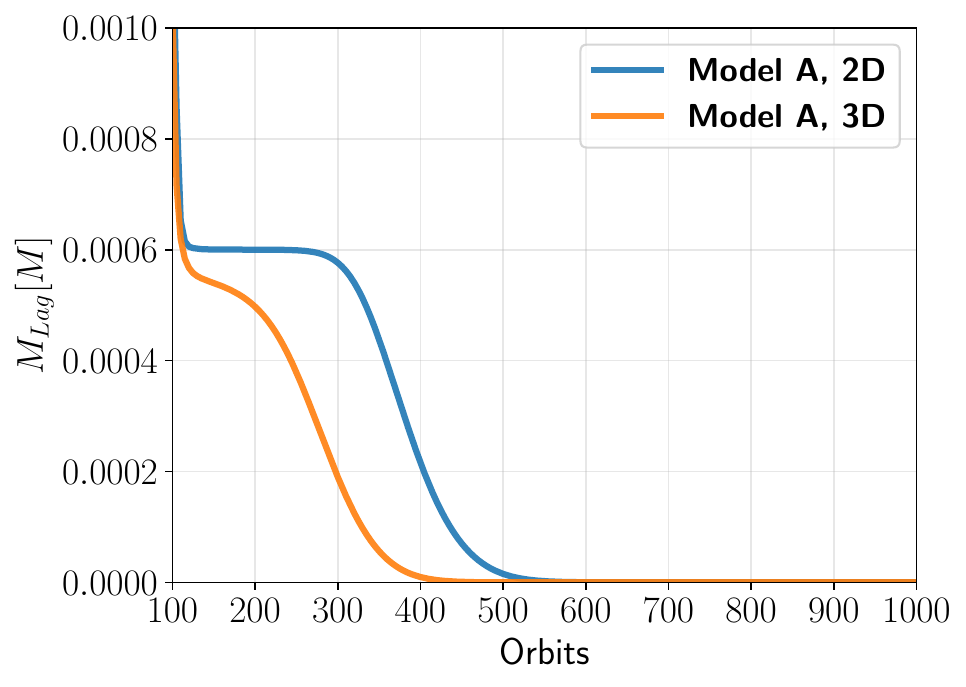}
        \caption{Dust mass inside Lagrangian traps in the 2D simulation (blue line) and 3D simulation (orange line) in Model A.}
        \label{figs:Lag_dustmass}
\end{figure}

In Model A, Lagrangian traps are inherently unstable and disappear after a few hundred orbits. 
Figure~\ref{figs:Lag_dustmass} shows the total mass of the dust trapped in Model C over time. 
The initial rapid decline after the release of dust at 100 orbits corresponds to the transition from the uniform initial dust distribution to the formation of Lagrangian traps.

In the 2D simulation, the Lagrangian trap remains nearly perfect for about 200 orbits following the dust release, after which it begins to leak rapidly and completely disappears by 400 orbits. 
In contrast, the 3D simulation shows significantly greater instability. 
While the Lagrangian trap forms quickly, it begins leaking almost immediately and vanishes entirely by 300 orbits. 
This enhanced leakage in the 3D simulation parallels the behavior of the pressure bump trap, but the trap’s duration is considerably shorter compared to the 2D case.

Several Lagrangian trap candidates have been proposed in observations, including those in HD 163296 \citep{Isella2018}, LkCa 15 \citep{Long22}, and PDS 70 b \citep{Balsalobre23}. 
Among these, only a single arc structure has been observed in HD 163296, whereas more than one arc structures have been detected in both LkCa 15 and PDS 70 b. 
Regarding planetary mass, PDS 70 b is a confirmed gas giant with a mass of approximately $3M_{\text{Jup}}$. 
\citet{Rodenkirch2021} suggested that a planet with at least $0.5M_{\text{Jup}}$ is required to produce the observed structure in HD 163296. 
For LkCa 15, simulations by \citet{Zhang2018} constrain the planetary mass to the range $0.1\text{–}0.3M_{\text{Jup}}$.

In our simulations, the planetary masses are relatively small ($0.03\text{–}0.3M_{\text{Jup}}$). 
Based on our test runs (not shown here), we find that as planetary mass increases, the formation of Lagrangian traps becomes more favorable, with notable changes in the asymmetry between L4 and L5 as well as the duration of the traps. 
Furthermore, we recognize that other parameters fixed in our simulations, such as the $\alpha$-parameter and the Stokes number, also influence the behavior of Lagrangian traps (e.g., \citealt{Lyra09}, \citealt{Zhang2018}, \citealt{Montesinos2020}).  
Lower $\alpha$-values ($\lesssim 10^{-4}$) promote the formation of long-lived vortices, enhancing dust trapping through aerodynamic drag. 
In contrast, higher $\alpha$ values ($\gtrsim 10^{-2}$) suppress these structures via turbulent diffusion, leading to weaker dust concentration.  
For the Stokes number, intermediate values ($\sim 0.1 - 1$) optimize trapping by balancing drag-induced migration toward pressure maxima with partial decoupling from gas flows, enabling rapid dust accumulation. 
Smaller Stokes numbers ($\ll 0.1$) result in particles remaining tightly coupled to the gas, delaying collapse, while larger values ($\gg 1$) cause grains to decouple entirely, escaping the traps before reaching critical densities.  
Thus, low $\alpha$ combined with moderate Stokes numbers maximizes the planet-forming potential of Lagrangian traps, while high $\alpha$ or extreme Stokes numbers inhibit their effectiveness.

Future work expanding the parameter space to include higher planetary masses and varying $\alpha$-parameters and Stokes numbers will be instrumental in determining the conditions necessary for L4/L5 trap formation, L4/L5 asymmetry, and their transient nature.

\subsection{Implication on Planet Formation and Disk Chemistry}

We now discuss consequences of leaky dust traps.
First, pebble isolation mass would not present perfect isolation; instead, it would act as a filtration (albeit still imperfect). \citet{Bitsch18} performed 3D single-fluid hydodynamic simulations to define pebble isolation mass as the mass of the core that would create a positive pressure gradient in its vicinity (i.e., the formation of a local pressure maximum; see their Figure 1; see also \citealt{LambrechtsJohansen2014HaltingPebble}). In our 3D, two-fluid simulations where we directly compute the dust motion, we find that such a definition is not sufficient to completely trap the pebbles within the planet-driven pressure bump, because the dust motion is mainly controlled by its coupling to the perturbed gas flow rather than the aerodynamic drift. Adopting the scaling relationship for the pebble isolation mass computed by \citet{Bitsch18}, our numerical setup with $\alpha=10^{-3}$ corresponds to $\sim$26$M_\oplus$, equivalently $q \sim 7.8\times 10^{-5}$ as the isolation mass (using the semi-analytic expression of the pebble isolation mass including dust diffusion by \citet{Ataiee18}, we arrive at the same mass). This mass is below that of our Model C and above that of our Model D so we expect a cumulative loss of dust of $\gtrsim$20\%. 

\citet{Bitsch18} also provide an expression for the minimum St of particles to be trapped when dust turbulent diffusion is taken into account (see their equations 17--22) under the premise that the dust motion is primarily driven by aerodynamic drag and turbulent diffusion. Directly computing the gas headwind velocity in our 3D simulations, we find that this minimum St ranges between 0.01 to 0.2 within the dust trap for models A \& C and ranges between 0.075 to 0.2 for Model B. For Model D, the minimum St stays flat at 0.01. While this is generally in alignment with stronger trapping in Models A \& C compared to Model B, the large fluctuation of this minimum St within the dust trap and the fact that our chosen St=0.1 is within the range yet still leaks suggests some incompleteness in the analytic theory. For instance, our simulations show that the dust motion is primarily driven by coupling to the gas flow so the base premise of the derivation of minimum St reported by \citet{Bitsch18} does not seem appropriate. A revision of analytic or semi-analytic trapping criteria would be welcome although a simple 1D analytic modelling appears challenging to us given the apparent lack of symmetry in the dust motion.

All of our simulations are computed at St = 0.1 for numerical efficiency; however, most of the grains that are trapped in the rings that we observe may have much smaller St $\sim 10^{-4}$--10$^{-2}$ \citep{Lee24}.\footnote{\citet{Lee24} derived the St and $\alpha$ relevant to observed dust rings assuming time-invariant trapping ratio of 0.6 and matching the dust mass locked within the ring to the initial mass reservoir in the outer disk under dust radial drift by coupling to the gas radial motion and aerodynamic drag. Lower trapping ratio would increase St since more dust would need to arrive at the location of the ring to match the measured ring dust mass. We defer a more quantitative calculation to a follow-up work.} Given that even for relatively high St particles, their dynamics through the trap is dominated by their coupling to the gas flow, we would expect even more leakage for smaller dust particles \citep{Zhu12}. Such a prospect implies the pebble isolation mass is not truly an isolation and its mass can continue to grow due to the solids that pass through \citep[e.g.,][]{Kalyaan2021, Kalyaan2023}.

Second, gap-opening planets do not perfectly prevent the exchange of material between the inner and the outer disk, suggesting that the formation of Jupiter alone may not explain the non-carbonaceous (NC) vs.~carbonaceous chondrite (CC) meteorite dichotomy in the solar system which is often interpreted as a clear spatial divide of volatile-poor vs.~volatile-laden solids \citep[e.g.,][]{Trinquier07,Warren11,Klein20}. 
Similar arguments against the role of Jupiter have been made in the literature based on 1D and 2D simulations finding particles initially trapped inside the pressure bump to grind down and the resultant small particles to more easily escape the trap \citep{Drazkowska2019,Stammler23}.

In general, some level of mixing between the NC and CC solids is expected and may better explain the continuous increase in isotopic anomalies with the inferred accretion age \citep[e.g.,][]{Onyett23,Homma24}. Such a temporal shift in the composition of solids in the solar system may be explained by an outward viscous spreading of the disk followed by a delayed delivery of the outer CC material by Jupiter \citep[e.g.,][]{Liu22} or thermal processing of the disk through energetic outburst events like FU Ori that initially pushes out the condensation fronts which then shrinks with time allowing for a gradual mixing \citep[e.g.,][]{Colmenares24}.\footnote{While this latter hypothesis has no need of Jupiter, Jupiter may nevertheless play a role in sculpting the overall solar system meteoritic composition including the heightened abundance of calcium-aluminum rich inclusion (CAI)s in CCs \citep[e.g.,][]{Desch18}.}

How much of the leaked dust pass by the planet and populate the inner disk? Within a smooth disk, the pebble accretion efficiency (defined as the rate of mass growth by pebble accretion over the rate of mass radial drift) is computed to be $\lesssim 10$\% \citep[e.g.,][]{Chachan23}. If the disk was smooth, we may therefore expect most of the leaked dust grains to flow interior to the planet's orbit rather than being accreted onto the planetary core. For gap-carving planets like we studied in this paper, the picture is more complicated as the dust grains flow with the gas following the spiral arms. Since this flow is more ``directed'' towards the planet, the relevant accretion efficiency onto the planet may be higher. \citet{Petrovic24} run similar multi-fluid 3D hydrodynamic simulations with more massive embedded planet ($q=10^{-3}$) and find that the fraction of the filtered dust that accrete onto the planet diminishes dramatically towards St $\gtrsim 10^{-3}$ owing to larger grains undergoing more rapid aerodynamic drift compared to smaller grains. Such a behavior is expected in smooth disks as well: for such high mass planet, pebble accretion is in the 2D, shear-dominated regime for which \citet{Chachan23} derive accretion efficiency $\propto {\rm St}^{-1/3}$ provided that $\alpha < {\rm St}$ (see their equation 11). 
In our simulations (which are not smooth disks owing to planetary perturbation), we find the dust motion over almost the entirety of our simulation domain to be governed by the coupling to the gas motion. Therefore, we may expect an opposite effect of higher St particles being more efficiently accreted by the planet once those particles leak out of the pressure trap. Following up in detail the motion of leaked dust particles incorporating solid accretion onto the embedded planet in our numerical setup is a subject of future work.\footnote{\citet{Petrovic24} speculate the motion of the dust grains to be governed by aerodynamic drag rather than coupling to the gas motion. Without more quantitative verification, it is difficult to ascertain whether that is truly the case. If it is true, their result diverges from our results where we find the dust motion to be governed by coupling to the gas flow, which may stem from our choice of more realistic thermodynamics ($\beta = 1$) instead of local isothermal equation of state.}

Finally, the existence of ring-like structures would not necessarily preclude the existence of volatile species in the inner disk.
Curiously, \citet{Perotti23} report an abundance of water in the inner disk of the PDS 70 system, in spite of the confirmed two giant planets in the outer orbit. This water could have been delivered while the giant cores were forming since the amount of solid mass that drift pass the core is about an order of magnitude larger than that accreted onto the core \citep[e.g.,][]{Lin18}. The detection of both gas and dust within the gap, albeit depleted \citep[e.g.,][]{Keppler19,Facchini21}, suggest there exists at least some level of ongoing leakage of material through the pressure bump since without continuous replenishment, the inner disk would have been evacuated of dust by their radial drift \citep{Pinilla24}. PDS 70 b \& c are massive enough that we expect these planets to generate a nearly perfect trap based on the results of our simulation at least for St = 0.1 and the results of \citet{Petrovic24} suggest that while the smaller grains more likely escape the dust trap, they are also more likely to be accreted by the planet. It may be that a significant portion of the inner disk volatiles originate from collisional processing of larger grains that arrived there during the formation of PDS 70 b \& c. In another case, volatiles including water are detected by JWST inside the 70 au cavity of the transitional disk SY Cha \citep{schwarz24}, suggesting imperfect dust trap as well. In a sample of 8 disks with ALMA-detected dust gaps narrower and weaker than the ones in transitional disks, \citet{gasman25} found the presence of a gap does not result in weak water emission, again suggesting gaps do not completely block inward dust drift, and the variations in the level of water emission may depend on the time at which the gaps appear \citep{Mah24}.

Post-processing the motion of grains from single-fluid 3D hydro simulations using fixed size grains (rather than fixed St; \citealt{Price25}), \citet{VanClepper25} arrive at similar results with smaller grains readily passing through the orbit of planets, especially for planets of lower mass. Their equivalent St at the location of the planet is very small $\lesssim$0.01 for their smallest (1 cm) grain and they report such grains to be trapped outside the planet's orbit only when the planet mass is beyond $\sim$50$M_\oplus$ (equivalently $\sim$0.4 thermal mass for their setup), which qualitatively agrees with our calculation (Model C vs.~Model D). While \citet{VanClepper25} find complete trapping of 1 cm grains, we still find some leakage, even at our significantly higher St. Such a difference may be from our more realistic treatment of gas thermodynamics and direct integration of dust motion in response to gas including dust feedback, which is expected to drive more instability \citep[e.g.,][]{Yang20,LiuBai2023}.

\section{Conclusion}~\label{sec:conclusion}

We ran two-fluid 2D and 3D hydrodynamic simulations, directly computing the motion of dust grains within a protoplanetary gas disk perturbed by an embedded planet. In both 2D and 3D simulations, dust grains are trapped within a pressure bump that forms exterior to the planet's orbit. These traps however are leaky with the degree of leakage increasing with higher level of turbulent diffusion (assumed the same between gas and dust diffusion) and with lower planet mass. Crucially, we find the traps to be significantly more leaky in 3D compared to 2D simulations. The reason for this more leakage in more realistic three-dimension is the complex geometry of the dust radial motion, showing both azimuthal and poloidal asymmetry with its dynamics dominated by the coupling to the gas radial motion which in turn is shaped by the waves generated by the planetary perturbation. 

Leaky dust traps have several consequences. First, the pebble isolation mass typically quoted in literature is not truly isolating and its mass is expected to grow as the dust grains continue to filter through. Second, unless the embedded planet is at least of Saturn mass and/or the disk is weakly turbulent ($\alpha \lesssim 10^{-3}$), the pressure trap created by the planet allows a metered passage of---rather than an impenetrable barrier to---the solid mass reservoir from the outer to the inner disk, suggesting such planets may not play as big a role as previously thought in creating compositional dichotomy within protoplanetary disks. 
Finally, dust traps being imperfect even for marginally coupled grains (St = 0.1) have significant implications for interpreting volatile compositions observed in the inner regions of disks hosting outer dust gaps \citep[e.g.,][]{Perotti23,schwarz24,gasman25}, opening up the possibility of significant amount of solid mass penetrating through the gaps, enriching the inner disk with heavy elements.

\vspace{5mm}

We thank the anonymous referee who provided a report that helped improve the manuscript. We further thank Edward Bergin, Bertram Bitsch, Steve Desch, Anders Johansen, Satoshi Okuzumi, and Ruth Murray-Clay for useful discussions. Many of these discussions took place at the KITP Edgeplanets program in 2025, supported in part by grant NSF PHY-2309135 to the Kavli Institute for Theoretical Physics (KITP).
P.H. was supported by NSERC ALLRP 577027-22. 
F.Y. acknowledge the Chinese Center of Advanced Science and Technology for hosting the Protoplanetary Disk and Planet Formation Summer School in 2024 where part of this work is conducted.
E.J.L. gratefully acknowledges support by NSERC, by FRQNT, by the Trottier Space Institute, and by the William Dawson Scholarship from McGill University.
R.D. acknowledges financial support provided by the NSERC through
a Discovery Grant. X.N.B. is supported by National Science Foundation of China under grant No. 12233004, 12325304, 12342501. This research was enabled in part by support provided by BC DRI Group through the Cedar cluster, Calcul Québec (\url{https://www.calculquebec.ca/}), Compute Ontario (\url{https://www.computeontario.ca/}) and the Digital Research Alliance of Canada (\url{https://alliancecan.ca/}).

\vspace{5mm}

\software{astropy~\citep{Robitaille2013astropy}, Athena++~\citep{Stone2020}}

\appendix
\counterwithin{figure}{section}
\counterwithin{equation}{section}

\section{Mesh Structure}\label{app:mesh}
\begin{figure}
\centering

\includegraphics[width=1.0\textwidth]{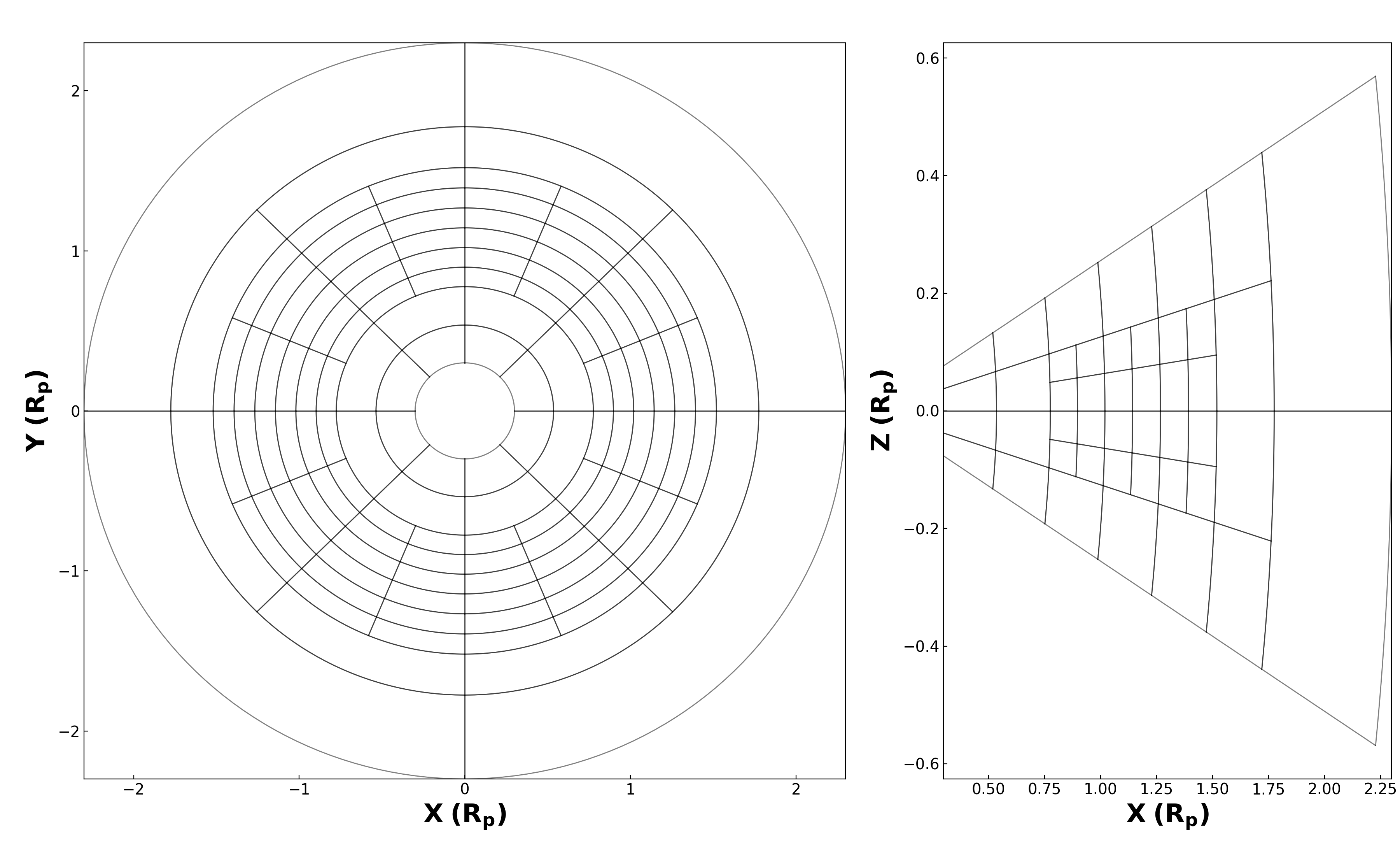}
\caption{The left panel displays the $X-Y$ section ($Z = 0$) of the mesh structure of 2D (3D) simulations, showcasing a grid size of $16\times48$ in one quadrangle on the $r-\phi$ plane. The right panel depicts the $X-Z$ section of the mesh structure for 3D simulations, with a grid size of $16\times16$ in one quadrangle on the $r-\theta$ plane.}
\label{fig:3D_meshblocks}
\end{figure}

In Athena++, the mesh structure is divided into units called meshblocks, each with the same grid sizes. In 2D simulations, each mesh block consists of $16 \times 48$ cells in the $R$-$\phi$ plane, while in 3D simulations, each meshblock contains $16 \times 16 \times 48$ cells in the $r$-$\theta$-$\phi$ space. Figure~\ref{fig:3D_meshblocks} illustrates the mesh structures used in the 2D and 3D simulations.

\section{The choice of edges}\label{app:boundary}

\begin{figure*}
    \centering
    \includegraphics[width=\textwidth]{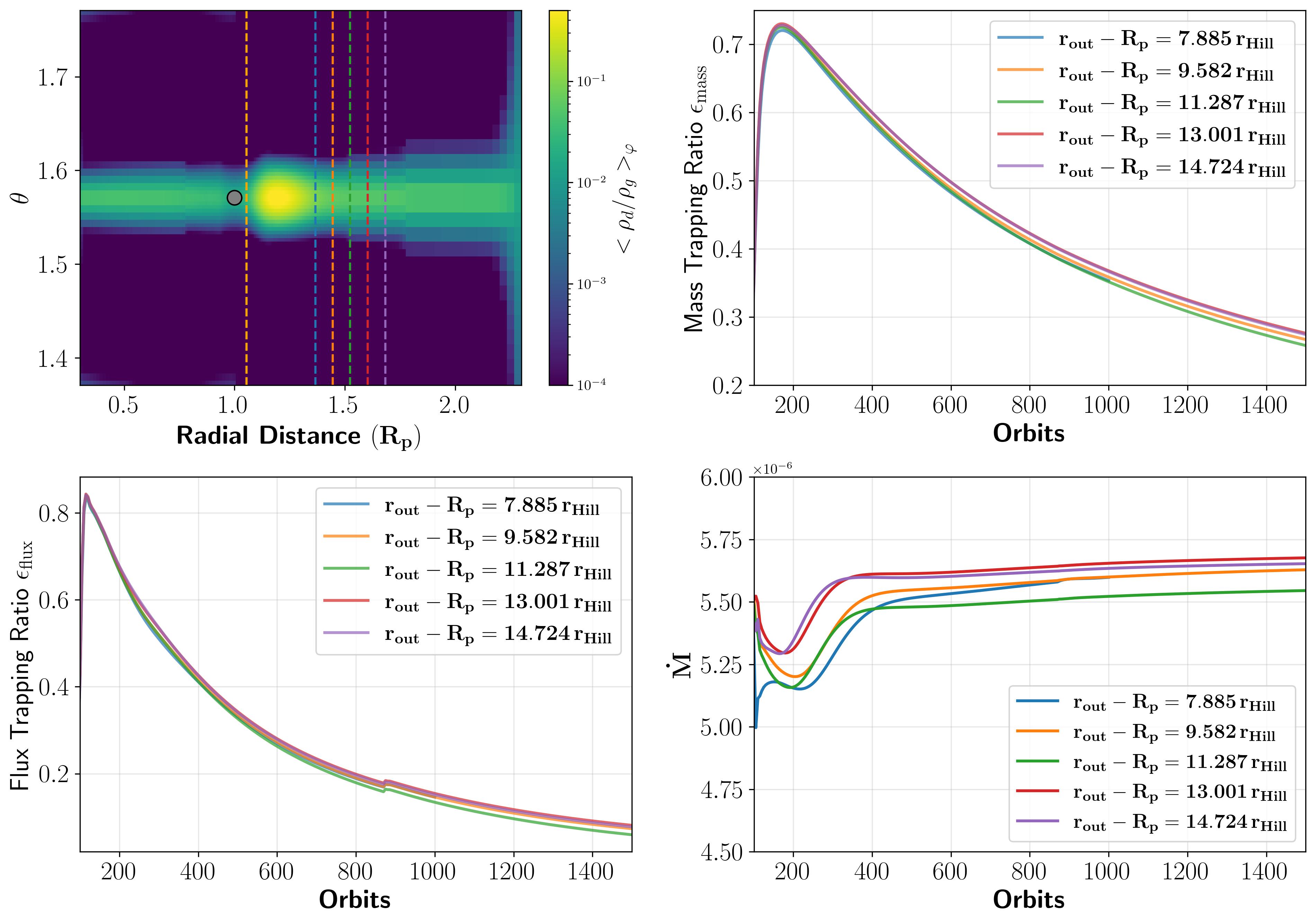}
    \caption{Effect of outer boundary selection on the mass flow through the outer boundary of the dust trap, the mass trapping ratio and the flux trapping ratio in 3D simulation Model B (see Table ~\ref{tab:Models}).
    \textbf{Top left:} $r-\theta$ distribution of the dust-to-gas ratio after azimuthal average. 
    The orange dotted line from the inside to the outside represents the inner boundary and the five different outer boundary values selected for the calculation. 
    \textbf{Top right:} Mass trapping ratio evolution.
    \textbf{Bottom left:} Flux trapping ratio evolution.
    \textbf{Bottom right:} Mass flux into the dust trap as a function of orbital time.
    }
    \label{fig:small_boundary}
\end{figure*}

We test the effect of outer boundary selection on trapping efficiencies $\epsilon_{\text{mass}} (t)$, $\epsilon_{\text{flux}} (t)$, and the dust mass flow $\dot{M}$ through the outer boundary. Figure \ref{fig:small_boundary} illustrates how there may be some variation in the mass flow (especially if the outer boundary is chosen too close to the dust trap), but the results for the mass trapping ratio and the flux trapping ratio are always robust with variations less than $\sim5\%$.

By default, $\theta$ boundaries are set to the edges of the computational domain in our 3D simulations where our boundary conditions enforce zero vertical dust flux. This setup facilitates a direct comparison with the 2D simulation.  We have verified that the dust layer remains concentrated within $\sim$0.2 of the gas disk scale height so that our $\dot{M}_{\rm ring}$ is robust to the changes in $\theta_{\rm min}$ and $\theta_{\rm max}$ as long as they stay beyond the scale height of the dust layer which is $\sim$0.1 of the gas disk scale height.

%\section[]{Galilean invariance in momentum diffusion}\label{app:galilean}
%\section{Explicit and Semi-implicit Drag Integrators}~\label{app:otherdrag}

\bibliographystyle{aasjournal}
\bibliography{references}{}

\end{document}